\documentclass[%
 preprint,
 amsmath,amssymb,
]{revtex4-1}

\usepackage{graphicx}
\usepackage{amssymb}
\usepackage{setspace}
\usepackage[utf8]{inputenc}
\usepackage{array}
\usepackage{natbib}
\usepackage{amsmath}
\usepackage{multirow}
\usepackage{hyperref}
\usepackage{rotating}
\usepackage{adjustbox}
\usepackage{threeparttable}
\usepackage{booktabs}

\usepackage{color}
\definecolor{redcolor}{rgb}{1.0,0.,0.}

\definecolor{bluecolor}{rgb}{0,0.,1}

\begin{document}

\preprint{}

\title{Machine learning and economic forecasting: the role of international trade networks}

\author{Thiago C. Silva$^1$, Paulo V. B. Wilhelm$^1$ and Diego R. Amancio$^2$}

\affiliation{$^1$Department of Economics, Catholic University of Bras\'ilia, Bras\'ilia, DF, Brazil \\
$^2$Institute of Mathematics and Computer Science, University of S\~ao Paulo, S\~ao Carlos, Brazil\\\\
}


\newpage 

\begin{abstract}
This study examines the effects of de-globalization trends on international trade networks and their role in improving forecasts for economic growth. Using section-level trade data from nearly 200 countries from 2010 to 2022, we identify significant shifts in the network topology driven by rising trade policy uncertainty. Our analysis highlights key global players through centrality rankings, with the United States, China, and Germany maintaining consistent dominance. Using a horse race of supervised regressors, we find that network topology descriptors evaluated from section-specific trade networks substantially enhance the quality of a country's GDP growth forecast. We also find that non-linear models, such as Random Forest, XGBoost, and LightGBM, outperform traditional linear models used in the economics literature. Using SHAP values to interpret these non-linear model's predictions, we find that about half of most important features originate from the network descriptors, underscoring their vital role in refining forecasts. Moreover, this study emphasizes the significance of recent economic performance, population growth, and the primary sector's influence in shaping economic growth predictions, offering novel insights into the intricacies of economic growth forecasting.
\end{abstract}

\maketitle



\section{Introduction}

In the evolving landscape of global economics, accurate forecasting of economic growth emerges as a cornerstone for crucial tasks, such as policymaking and investment decisions~\cite{elliott2008economic}. Traditional models for predicting GDP growth heavily rely on traditional economic indicators and standard linear econometric techniques~\cite{de200625}. However, the increasing complexity of global trade and economic interdependence calls for a more nuanced approach to forecasting. Recent advances in machine learning offer promising avenues for enhancing the performance and reliability of economic forecasts by processing large datasets and identifying complex, non-linear relationships~\cite{masini2023machine}. This study stands at two main strains of literature: analyzing international trade networks and applying machine learning models in economic forecasting. Specifically, it innovates by utilizing topological measures of international trade networks as features in machine learning models to forecast GDP growth and showing that they substantially enhance a country's GDP growth prediction. 

The significance of our work lies in its potential to provide a more comprehensive understanding of the factors driving economic growth predictions. This is particularly relevant in a world where economic landscapes are rapidly changing, and traditional forecasting models often fail to capture the dynamism of global interactions shaped by economic flow between countries as world trade. By integrating international trade network topologies into the machine learning framework, this study opens new pathways for research and offers practical insights for economists, policymakers, and investors seeking to navigate the complexities of the global market.

A substantial corpus of research leverages complex network theory to shed light on many topics in economics and finance~\footnote{This body of work is extensive and includes analyses such as the behavior of stock returns amidst the Global Financial Crisis~\cite{silva2023effect}, the fluctuation patterns in stock prices~\cite{chi2010network}, and the intricate relationship between the advent of COVID-19 and stock market dynamics~\cite{samitas2022covid}. Further exploration within this field includes the examination of the financial impacts stemming from shifts in monetary policy~\cite{silva2020micro} and the formulation of investment strategies within a global framework~\cite{lee2019global}. The literature also navigates through the interconnectedness of global banking networks~\cite{demirer2018estimating}, the correlation between a firm's position in the supply chain structure and its innovation capabilities~\cite{chuluun2017firm},  and the evaluation of financial system vulnerabilities and the spread of economic impacts~\cite{silva2017monitoring}.}. The first part of our work deals with constructing trade networks and extracting topological measures.  Multiple approaches to modeling the International Trade Network (ITN) are documented, such as binary and weighted configurations or directed and undirected networks~\cite{garlaschelli2004fitness,barthelemy2005characterization}. Additionally, these models vary in terms of granularity, ranging 
from aggregated overviews to detailed breakdowns by very specific commodity types. Our research focus on the domain of commodity-specific trade networks, aligning with notable studies that explore various dimensions of the ITN through the lens of particular commodities~\footnote{These studies encompass a wide range of topics, including the evolutionary dynamics of the international fossil energy trade's multilayer network~\cite{gao2015features}, the pivotal roles countries play within the international fossil fuel trade network~\cite{zhong2017roles}, the spatiotemporal evolution of global plastic waste trade networks~\cite{wang2020structure}, and the community structure of the food-trade international multi-network~\cite{torreggiani2018identifying}. Further contributions to the field examine the spatial-temporal evolution of the global copper raw materials and scrap trade networks~\cite{hu2020characteristics}, the characteristics and stability of the global complex nickel ore trade network~\cite{ma2022analysis}, the analysis of wood and non-wood forest global trade network~\cite{lovric2018social}, the temporal stability of international fertilizer trade networks, and the multiplex network structure of global cobalt industry chain~\cite{shi2022multiplex}.}.

Our contribution to this literature has two principal dimensions. Firstly, we endeavor to explore commodity-specific trade networks through a section-level categorization as delineated by the Harmonized System (HS) code framework. This approach enables us to discern nuances in the topological analysis that are particular to this level of granularity, nuances that are frequently overlooked in current literature~\footnote{Prevailing studies on international trade networks predominantly utilize aggregate trade data or examine it at a highly granular level. Adopting a section-level perspective for trade data affords us a compromise between these two extremes, which is crucial since the granularity of data can significantly influence topological characteristics.}. Secondly, we concentrate on the temporal evolution of the topological properties of the most significant commodity-specific networks from 2010 to 2022. This timeframe is especially pertinent for investigating trends of de-globalization, which have been intensified by critical global events such as the COVID-19 pandemic and the geopolitical tensions following Russia's invasion of Ukraine. Through this perspective, our analysis aims to elucidate the complexities of how commodity trade networks have adapted during these tumultuous periods, identifying which countries have either derived benefit from or been adversely affected.

In our analysis of the temporal dynamics of assortativity and network density, we observe a critical juncture between 2016 and 2018 that marks a reversal of previously noted trends, leading to significant topological transformations within the main section-level trade networks. This period coincides with a marked increase in trade policy uncertainty, underscoring the profound impact that trends towards de-globalization exerted during this timeframe. Furthermore, we develop centrality rankings for the main commodity trade networks for the years 2010 and 2022, enabling the identification of key global actors and notable shifts in their influence over this period. Our findings affirm the United States' continued dominance as a leading entity across four of the five principal commodity trade networks in 2022. Similarly, China and Germany have sustained high performance, consistently appearing in the top three or five rankings. This research documents various nations' ascension and decline in these rankings. Notably, Thailand's presence diminished significantly, falling from being in the top fifteen in 2010 to its absence in 2022 across the five main trade networks. In contrast, India showcases a remarkable trajectory, not featured in any of the top fifteen rankings in 2010, yet in 2022, it emerged in all, including three instances within the top five.

The second segment of our study engages with the literature that applies machine learning techniques to forecast economic growth. An expanding collection of research highlights that machine learning methods can surpass traditional econometric models in predicting GDP growth~\citep{chuku2019intelligent, richardson2021nowcasting, vrontos2021modeling, goulet2022machine}. While these studies incorporate a diverse array of predictors, to our knowledge, they still need to harness network topology metrics from international trade networks as predictive features~\footnote{Several works utilize complex networks in forecasting tasks, including the use of stock networks to predict economic growth~\cite{hegre2010trade} and the predictability of stock price series through community detection and network metrics~\cite{anghinoni2019time}.}. We argue that the integration of topological metrics into machine learning models can markedly enhance the precision of economic growth forecasts. Trade network-based descriptors are pivotal for two reasons. First, a country's trade network position reflects its global significance and product/service exchange flexibility. Second, economic trends in neighboring countries can indicate a nation's own economic prospects, such as GDP growth challenges highlighted by reduced transactions with a neighbor. Thus, incorporating trade network measures enriches predictive models with a comprehensive view of global economic interactions, enabling us to identify patterns and trends that traditional economic indicators may overlook.

Our study contributes by assessing the importance of network measures from commodity-specific trade networks on the performance of machine learning's economic growth forecast. A distinctive feature of the literature is that it generally focuses on one or a moderate amount of countries. In this sense, we contribute by employing a "horse race" of machine learning models to find which performs best in forecasting economic growth for a comprehensive set of more than 200 countries.\footnote{Refers to a comparative evaluation among a variety of machine learning models to determine which performs best on a given task or dataset. This analogy draws from actual horse races, where horses compete to see which is fastest under similar conditions to identify the top performer.} Accurate economic growth forecasts are essential as they guide strategic decisions and policy development globally. By enabling policymakers to preempt economic instabilities, these forecasts contribute to a more stable economic climate. For businesses and investors, such insights are crucial for navigating market uncertainties and seizing growth opportunities. Additionally, in a closely interconnected global economy, the accuracy of these predictions is vital for international coordination in addressing economic challenges like trade imbalances and inflation. Thus, precise forecasts play a pivotal role in both national and global economic stability and collaboration.

In establishing a foundation for our analytical framework, we employ the performance of the linear regression model as a baseline. This technique, notable for its simplicity in implementation and clarity in model interpretation, is a benchmark against which the efficacy of more advanced algorithms can be measured. By doing so, we facilitate an examination of the trade-offs inherent between model complexity and performance enhancement. To surpass the established baseline, we introduce a suite of sophisticated models into our study, including the Support Vector Machine, Light Gradient Boosting Machine, k-nearest Neighbor, Random Forest, and XGBoost.

Given that these advanced methodologies may compromise the interpretability afforded by linear regression, we employ the SHAP value analysis—a tool derived from the concept of Shapley value borrowed from game theory. This analytical tool enables us to restore a level of interpretability to these more complex and performant models. Through this approach, our contribution to the literature extends beyond merely identifying the superior predictive models. We provide a detailed exploration of the specific features that exert the most significant influence on economic growth forecasts and elucidate the mechanisms through which they operate. Consequently, our work enriches the literature with a comparative analysis of model performance and a deeper insight into the dynamics that drive economic growth predictions.

In our empirical evaluation, we deploy four distinct error metrics to ascertain the average error values, thereby gauging the performance of models. Our findings delineate that the Random Forest, XGBoost, and k-NN models, in respective order, emerge as the superior performers, showcasing substantial enhancements over the baseline model. These outcomes suggest that the aforementioned models possess an inherent capacity to navigate the complexities and non-linearities within the dataset, a task at which a simplistic linear regression model proves inadequate.

Further examining feature importance within the three most efficacious models yields noteworthy insights. Primarily, we observe that nearly half of the top fifteen most influential features for the Random Forest model—the most effective one—are network measures. This underscores the pivotal role of commodity trade network topologies in refining economic growth forecasts. Particularly, the density of the Mineral trade network stands out as a critical feature, ranking as the second most significant for both Random Forest and XGBoost models and as the foremost for the LightGBM.

Additionally, our analysis brings to light the autoregressive nature of economic growth, where the immediacy and recency of GDP growth figures are fundamentally pivotal in forecasting economic performance. This observation lends credence to the concept of "economic inertia," wherein past performance exerts a lingering influence on future outcomes. Moreover, we identify the significance of the modularity of the Machine \& Electrical trade network, population growth, and relevance of the primary sector as key predictors of economic growth. These elements consistently feature among the top fifteen most relevant features across all three top-performing models, reinforcing their importance in the predictive modeling of economic performance.

To further elucidate interpretability insights, we conduct a comprehensive feature interpretability analysis on the most effective model, the Random Forest. In accomplishing this task, we utilize a suite of Shapley value dependence plots, which furnish a detailed perspective on how fluctuations in feature values impact economic growth forecasts. One particularly intriguing pattern emerges from analyzing the Mineral trade network's density. We observe moderate escalations in this network measure are associated with neutral or beneficial effects on economic growth predictions. However, once a critical threshold is exceeded, any additional increase in network density is linked to a decline in future economic prospects. This discovery suggests an optimal range within which enhancements to network density can positively influence future economic growth, marking a critical insight into the nuanced interplay between trade network characteristics and economic forecasting.

Our research elucidates several policy implications of paramount importance. Firstly, the significant topological transformations we document within global trade networks highlight their vulnerability to de-globalization trends. This observation necessitates a strategic reassessment by policymakers of economic policies heavily reliant on globalization mechanisms. In tandem, international collaboration must be fortified by a commitment to transparency in the formulation and implementation of policies. Such an approach is essential in mitigating uncertainty and cultivating a stable milieu for international trade.

Furthermore, our analysis of centrality rankings unveils an evolving paradigm within the domain of international trade. This evolving landscape necessitates a proactive and visionary approach to policy formulation. The sustained dominance of established economies, juxtaposed with the meteoric rise of emerging markets like India, signals dynamic shifts propelled by technological advancements and geopolitical developments.

Moreover, the outcomes of our comparative analysis of forecasting models reveal the intricate and non-linear nature of economic growth predictors. This complexity underlines policymakers' need to employ more sophisticated analytical frameworks in their forecasting endeavors. Importantly, integrating network measures into forecasting models emerges as a powerful strategy to amplify the accuracy of economic growth predictions. Such enhanced predictive capability is invaluable, furnishing policymakers with the insights to craft more informed and efficacious policy measures.

\section{Network analysis of section-level international trade networks}
\label{sec:net_anal}

Our network analysis framework is structured into three subsections. Initially, Subsection~\ref{subssec_netdata} furnishes a data description and details the database utilized to construct section-level trade networks. Subsequently, Subsection~\ref{subssec_netmeas} delineates the network measures employed to examine the topological properties of commodity-specific trade networks. Finally, in Subsection~\ref{subssec_topanal}, we explore the temporal dynamics of five fundamental network measures spanning 2010 to 2022. This section also analyzes shifts in centrality rankings, offering insights into the evolving structure and influence dynamics within these networks over time.

\subsection{Data description}
\label{subssec_netdata}

Our research enhances the understanding of global trade by constructing networks derived from detailed trade data. Utilizing the Comtrade database provided by the United Nations Statistics Division, we analyze bilateral merchandise trade information for 261 countries, dependent territories, and special areas of geographical interest.\footnote{We follow three-letter country codes defined in ISO 3166-1, part of the ISO 3166 standard published by the International Organization for Standardization (ISO).} This database organizes trade data according to the Harmonized System (HS) classification.\footnote{
The Harmonized System of tariff nomenclatures, often referred to as the Harmonized System Classification, is a standardized system of names and numbers for classifying traded products. Developed and maintained by the World Customs Organization (WCO), the HS is utilized by over 200 countries and economies as a basis for their customs tariffs and for collecting international trade statistics. The HS is designed to facilitate the international trade process by simplifying the identification and classification of goods across borders. The system is structured into 21 sections, further divided into 97 chapters. These chapters are subdivided into headings and subheadings, representing a hierarchical structure that classifies goods from the general to the specific. For more information, access: \href{https://www.wcoomd.org/en/topics/nomenclature/instrument-and-tools/hs-nomenclature-2022-edition/hs-nomenclature-2022-edition.aspx}{HS Nomenclature 2022 edition}.} Our approach involves extracting data at the chapter level and then aggregating this information into section level, providing the most general categorization. Given the extensive nature of the official names for these sections, we use shortened names to improve both brevity and clarity in our presentation. Table~\ref{tab:section_names} lists the code and associated abbreviated name for the ten most significant sections in terms of trade flow value from 2010 to 2022. The top five sections during this period accounted for approximately 60.7\% of the global trade flow value.

Our data has a monthly frequency, which is subsequently consolidated to accommodate analyses on both quarterly and annual scales, contingent upon the specific context of the study. Transactions classified as re-importations and re-exportations are excluded to concentrate on standard trade activities. The Comtrade database records each trade occurrence from the perspectives of both participating entities (the importer and the exporter), thereby introducing a potential mismatch in the transacted volume between each pair of reporting entities. To mitigate the inconsistencies arising from this dual reporting, this study adopts a methodology where the data reported by the nation exhibiting a superior aggregate trade value within the network is prioritized. This approach is grounded in the hypothesis that data furnished by nations with larger trade values tend to be more accurate, particularly when trade involves nations with significantly disparate economic sizes.

\begin{table*}[!ht]
\centering 
\begin{threeparttable}
\caption{The Top 10 most relevant commodity sections from 2010 to 2022.}
\label{tab:section_names}
\begin{tabular}{ccc}
\toprule\midrule
Code & Abbreviated name & Relevance \\
\hline
16 & Mechanical \& Electrical & $24.3\%$ \\

5 & Mineral & $15.1\%$ \\

17 & Transport & $10.5\%$ \\

6 & Chemical & $10.0\%$ \\

15 & Base Metals & $7.2\%$ \\

7 & Plastics \& Rubber & $4.6\%$ \\

11 & Textile & $4.2\%$ \\

14 & Precious Metals & $3.9\%$ \\

18 & Instruments & $3.6\%$ \\

4 & Beverages \& Tobacco & $3.4\%$ \\

\midrule\toprule
\end{tabular}

\begin{tablenotes}
\footnotesize
  \item[1] In this table, "Code" corresponds to the identifier assigned to each commodity section according to the Harmonized System (HS) classification. "Abbreviated name" denotes the concise name we have devised to represent the official title of the section as defined by the HS classification. "Relevance" represents the section's trade value as a percentage of the total merchandise trade value from 2010 to 2022.
\end{tablenotes}

\end{threeparttable}

\end{table*}

\subsection{Relevant background: network measurements from the complex networks literature}
\label{subssec_netmeas}

In the commodity-specific trade networks we construct, nodes represent individual countries, while edges denote bilateral trade flows of a particular commodity section. The edge is directed and goes from the exporting (or "origin") country to the importing (or "destination") country. The weight of each edge signifies the monetary value of the respective bilateral trade flow. A comprehensive array of metrics for describing network topology is available in existing literature. For our analytical purposes, we adopt a set of network metrics as delineated by~\cite{silva2016machine}, encompassing a diverse range of fundamental measures that encapsulate local, intermediate, and global network attributes. This strategic selection enables us to elucidate the multifaceted topological characteristics of each section-level international trade network, ranging from the nuanced intricacies of individual trade connections to the overarching structure of global trade within the specific commodity category.

 \vspace{5mm}

\begin{itemize}
    \item \textit{Strength (strictly local):} defined by the aggregate weight of edges connected to a node, node strength in directed graphs splits into in-strength and out-strength. In-strength calculates the total weight of incoming edges to a node, analogous to a country's total import value within a specific trade section. Conversely, out-strength tallies the weight of outgoing edges, akin to a country's total export value. 
    
     \vspace{5mm}
    
    \item \textit{PageRank (mixed):} by leveraging the PageRank algorithm, we determine the relative importance or centrality of countries within the trade network. A country's centrality reflects not just its direct trade connections but also the significance of its trading partners.   The methodology for our PageRank calculation emphasizes incoming edges—interpreted, within the framework of our analysis, as import flows. Consequently, our centrality measure is designed to recognize a country's influence within the network based on import performance.
    
    \vspace{5mm}
    
    \item \textit{Clustering coefficient or Transitivity (mixed):} this metric evaluates the degree to which nodes in the network cluster together, calculated by the ratio of actual triangles to potential triangles among nodes. High clustering coefficients (close to 1) indicate a high level of local clustering, while low values (close to 0) suggest a more dispersed trade network.

      \vspace{5mm}
    
    \item \textit{Density (global):} provides a snapshot of the fulfillment of the network's connectivity potential, measured by the ratio of actual to possible connections. A higher proportion of potential trade links are realized in dense trade networks, indicating a closely interconnected global trade fabric.

     \vspace{5mm}
    
    \item \textit{Assortativity (global):} evaluates the degree to which edges in the network join nodes sharing analogous attributes. We employ the assortativity degree coefficient, which measures the frequency of connections between nodes with similar degrees. A value close to 1 indicates an assortative network, suggesting that many trade connections occur between countries with similar numbers of trading partners, whether they are many or few. On the other hand, a value close to -1 points to a disassortative network. This reflects a pattern where trade connections frequently exist between countries with disparate numbers of trading partners—one with a high number of partners and the other with a low number. This approach clarifies the nature of connectivity within the international trade network, revealing the tendencies for countries to trade with partners of similar or dissimilar connectivity levels.

     \vspace{5mm}
    
    \item \textit{Reciprocity (global):} calculates the extent of mutual exchanges within the network based on the proportion of reciprocated connections relative to the total number of connections. A high reciprocity level, approaching 1, indicates that most trade partnerships involve mutual exchanges. Conversely, a low reciprocity level, nearing 0, suggests that the bulk of trade flows occurs in one direction between trade partners. 

     \vspace{5mm}
    
    \item \textit{Modularity (global):} explores how the network is segmented into distinct communities or modules, characterized by dense connections within each community and fewer connections between different communities. High modularity values, approaching 1, signal that connections within communities significantly outnumber those between communities. Conversely, values near 0 indicate a balance between connections within communities and those bridging different communities. This evaluation highlights the network's structure, identifying whether it is more segregated into tightly-knit groups or integrated, with fluid interactions across different groups.

     \vspace{5mm}
\end{itemize}

By applying these network measures, we aim to unravel the complex interdependencies and structural patterns characterizing section-level international trade networks, offering novel insights into the global trade landscape.

\subsection{Topological analysis of section-level international trade networks}
\label{subssec_topanal}

In this subsection, we provide a topological examination of section-level international trade networks, focusing on the period from 2010 to 2022. This interval is convenient for understanding the dynamics of globalization, as it marks the emergence of signs pointing to a deglobalization process. During this period, several pivotal events exerted a profound impact on global trade dynamics, including Brexit, the China-USA trade war, the COVID-19 pandemic, and Russia's invasion of Ukraine. Given the expansive scope of this analysis, our investigation is streamlined to concentrate on the five most significant commodity sections of this period. Collectively, these sections account for approximately 60\% of the monetary trade value within the specified timeframe, thereby providing a comprehensive overview of the critical trends and shifts in international trade patterns during a period marked by considerable geopolitical and economic upheavals.

\begin{figure}[]
  \centering
    \includegraphics[width=0.98\textwidth]{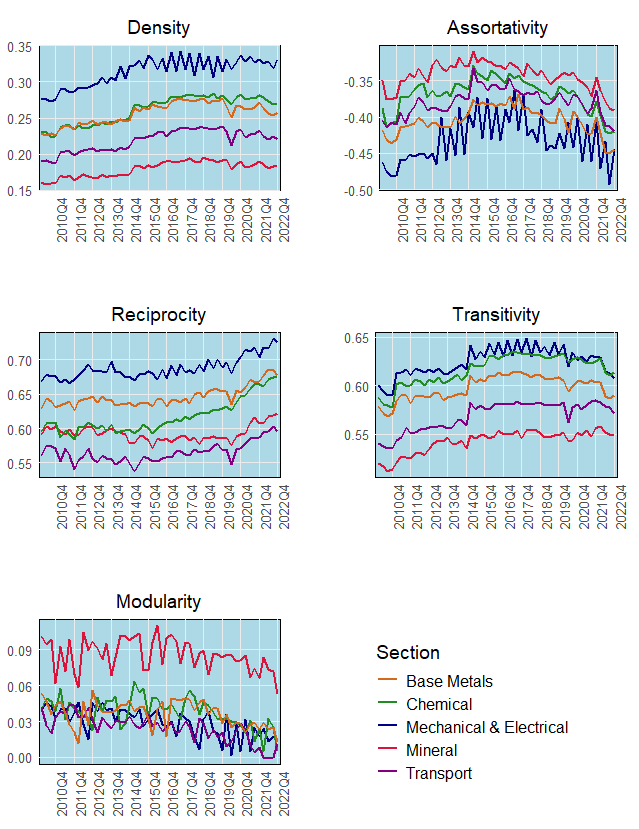}
  \caption{This figure presents an analytical overview through five distinct plots, illustrating the evolution of key topological metrics within international trade networks from 2010 to 2022. Each plot tracks the temporal progression of a particular network measure for each of the five main section-level trade networks under scrutiny. The analysis employs a quarterly temporal resolution.
}
  \label{fig:net_meas}
\end{figure}

Figure~\ref{fig:net_meas} depicts the evolution of five fundamental network measures for the most traded commodity sections from 2010 to 2022. Initially, we examine the network density measure. Among the sections analyzed, Mechanical \& Electrical exhibits the highest average network density value of 0.314, indicating that, on average, 31.4\% of the potential trade connections between countries are realized within this section. Conversely, the Mineral section presents the lowest average density value, 0.179, which corresponds to merely 57\% of the value observed for the Mechanical \& Electrical section. The average density values for the remaining sections are as follows: Chemical (0.261), Base Metals (0.256), and Transport (0.219). A noteworthy observation is the bifurcation in trends during the analyzed period; from 2010 to 2017, an increase in density values suggests enhanced interconnectedness among trading partners. However, post-2017, a trend toward stagnation or decline is observed. Notably, the Mechanical \& Electrical section demonstrates stability in its density value throughout the period, whereas the Base Metals, Transport, and Mineral sections exhibit a clear downward trajectory. The onset of the COVID-19 pandemic in early 2020 prompted a sharp decline in the density values of the Transport and Chemical sections, while the Mechanical \& Electrical section experienced a less pronounced impact, highlighting the heterogeneous effects of the pandemic on trade network interconnectedness.

Turning our attention to the assortativity network measure, all sections reveal predominantly negative assortativity values, suggesting prevalent trade interactions between heterogeneous countries (i.e., nations with a disparate number of partners). The Mechanical \& Electrical section stands out with the most negative average value of -0.432, while the Mineral section records the least negative average assortativity value of -0.343. The assortativity values for the other sections are as follows: Base Metals (-0.405), Transport (-0.379), and Chemical (-0.369). Temporal analysis of these values reveals two distinct phases. Until mid-2014, a shift towards less negative assortativity values indicates a trend towards more homogeneous commercial relationships. Contrarily, after 2017, this trend reversed, with assortativity values becoming more negative, culminating in 2022 with lower values than those recorded in 2010 for all sections, excluding Mechanical \& Electrical. This recent trend underscores the disassortative nature of these trade networks.

The analysis of the temporal evolution of assortativity and network density suggests that a pivotal shift occurred during 2016-2018, reversing earlier observed trends and inducing profound topological alterations within the main section-level trade networks. This period aligns with a significant surge in the Trade Policy Uncertainty Index~\cite{caldara2020economic}~\footnote{Originating in 1960, this index experienced its most pronounced increase during this timeframe. Specifically, the index's monthly average escalated from 27.3 (2010-2015) to 114.3 (2016-2019), with peaks exceeding 200 in certain months.}. Such dramatic rises in trade uncertainty underscore the detrimental impacts of critical geopolitical and economic events on global trade within the last decade. This correlation between the surge in trade policy uncertainty and our findings suggests that heightened trade policy uncertainty correlated with lower network connectivity and prompted more peripheral countries to re-establish connections with central trade hubs. The economic rationale behind this observed trend is attributed to an increased risk aversion amid rising economic uncertainty. In the period preceding 2016-2018, characterized by relatively low trade policy uncertainty, countries were more inclined to engage in international trade, actively seeking new partnerships, often with peripheral nations within the trade network. This era witnessed an increase in assortativity, with peripheral countries predominantly engaging in transactions with similarly positioned entities, i.e., other peripheral countries.

However, the landscape shifted dramatically after 2018 amidst the mounting trade uncertainty and heightened risk aversion. Trade patterns began to coalesce around more established, central countries within the network, denoting a strategic pivot in economic interactions. Peripheral nations, in response, scaled back their trade transactions with similarly peripheral counterparts, opting instead to prioritize transactions with more central entities, perceived as safer counterparts. This strategic realignment precipitated a decline in assortativity, marking a transformative phase in the connectivity patterns of the main section-level trade network.

Continuing our topological examination, we analyze the dynamics of reciprocity and clustering patterns across the primary commodity trade networks. Regarding reciprocity, all sections consistently exhibit average values above 0.500 throughout the analysis period, underscoring a prevalent trend of reciprocal trade relations among countries. The Mechanical \& Electrical section stands out with the highest average reciprocity, achieving 0.688, while the Transport section presents the lowest, at 0.565. The average reciprocity for the remaining sections is as follows: Base Metals (0.646), Chemical (0.617), and Mineral (0.592). A temporal analysis reveals distinct trends: the Mechanical \& Electrical section shows a unique and consistent increase in reciprocity throughout the period. The Base Metals and Chemical sections display relatively stable values up to 2015, followed by a noticeable rise. The Mineral and Transport sections initially exhibit a decline, yet from 2016, Transport experienced a resurgence in reciprocity, with Mineral showing recovery only from 2020 onwards. Interestingly, the Chemical and Mineral sections, which had similar reciprocity levels in 2010, diverged post-2014, with the Chemical's increased reciprocity aligning it more closely with Base Metals. By the end of the analysis period, all sections show a trend towards higher reciprocity, indicating a strengthening of reciprocal trade relations over time.

Concerning local clustering behavior, we note that each section maintains an average transitivity value above 0.540, indicating significant local clustering. The Mechanical \& Electrical section has the highest average transitivity at 0.624, with Mineral recording the lowest at 0.543. The figures for the other sections are Chemical (0.616), Base Metals (0.600), and Transport (0.569). Temporal dynamics of this metric unveil varying patterns: all sections witness an increase in local clustering up to mid-2014. From 2014 to 2019, transitivity stabilizes, halting the previous upward trend. The onset of the COVID-19 pandemic leads to a notable drop in clustering, especially within the Transport and Mechanical \& Electrical sections. After 2019, a general decline in clustering was observed, except for the Mineral section, which showed resilience by maintaining its pre-pandemic transitivity values. Additionally, the period under analysis sees the Chemical section's transitivity converging with that of the Mechanical \& Electrical, with the first ultimately exhibiting the highest transitivity by 2022.

We assess modularity patterns to conclude the first part of our topological analysis. Across all sections, average modularity values remain exceedingly low, approaching zero, throughout the studied period. Considering the significant local clustering identified earlier, these minimal modularity values highlight a key aspect of the networks' structure. Despite a tendency for countries' trade partners to form interconnected clusters, these section-level trade networks lack a modular organization, suggesting the absence of distinct segregation in trade interactions among specific groups. This finding suggests the existence of many countries serving as bridges connecting peripheral countries.

In the second part of our topological analysis, we explore shifts in ranking centrality within the primary section-level trade networks from 2010 to 2022. As detailed in Table X, our focus centers on the Mechanical \& Electrical, Chemical, and Mineral sections. Commencing with the Mechanical \& Electrical sector, we observe that the United States maintains its position as the most central nation throughout the period under review. China and Germany consistently rank within the top three, with Germany advancing to second place by 2022. Notably, Great Britain, Singapore, and Japan experienced a decline in their rankings, with Great Britain falling from fifth to tenth, Singapore from ninth to twelfth, and Japan from twelfth to fifteenth. Conversely, Mexico demonstrates significant progress, moving from seventh to fourth position. The dynamics of entry and exit from the Top 15 list highlight substantial shifts, with Thailand disappearing from the 2022 rankings after being fourth in 2010, and Russia, Malaysia, and Spain also absent in the later rankings. Conversely, Hong Kong, India, the Republic of Korea, and the United Arab Emirates emerged in the 2022 rankings, with Hong Kong and India notably securing seventh and ninth positions, respectively.

In the Chemical sector, the United States leads, followed by China and Germany, mirroring the outcomes observed in the Mechanical \& Electrical section. This period witnessed an increased prominence of the United States and China, with a noticeable gap emerging between them and the other nations ranked from third to fifteenth in 2022. The upward mobility of the Netherlands, Belgium, Canada, Brazil, India, and the Republic of Korea is remarkable, with India's rise from fourteenth to fourth and Brazil's from thirteenth to seventh being particularly noteworthy. Thailand's absence in the 2022 rankings, after being fifth in 2010, marks a notable decline. Additionally, the rankings reflect a downturn for France, Great Britain, Japan, Italy, and Spain, with the significant repositioning of Great Britain from sixth to tenth and Japan from seventh to eleventh.

For the Mineral section, China's dominance is unmistakable. The United States places second, even though its relative centrality to China decreased significantly from 80\% in 2010 to 45\% in 2022. This underscores China's consolidating dominance. Australia, the Republic of Korea, France, and the Netherlands see noteworthy advancements in the 2022 rankings, with the Republic of Korea's ascent from eleventh to seventh, highlighting its increased centrality. Conversely, Belgium, Hong Kong, Italy, Japan, and Singapore register declines, with Hong Kong's fall to fourteenth and Belgium's to fifteenth marking significant shifts. The rankings note the departure of only two countries, Thailand and Canada, from the Top 15 in 2010 and the entry of India and Malaysia in 2022, with India's placement at fourth being particularly impressive.

The second part of our topological analysis looks into shifts in ranking centrality within the main section-level trade networks between 2010 and 2022. Table~\ref{tab:centrality1} reports centrality rankings for the following sections: Mechanical \& Electrical, Chemical, and Mineral. Regarding the Mechanical \& Electrical section, the United States is the most central country for both years under review. Following closely, China and Germany constitute the Top 3, with Germany ascending to the second position by 2022. We highlight the fall in the rank position of the following three countries: Great Britain descends from the fifth to the tenth position, Singapore from the ninth to the twelfth, and Japan from the twelfth to the fifteenth. Conversely, Mexico exhibits an upward trajectory, advancing from the seventh to the fourth position. Additionally, entry and exit dynamics from the Top 15 list reveal notable shifts. Thailand, previously ranked fourth in 2010, vanishes from the 2022 ranking. Similarly, Russia, Malaysia, and Spain, present in the 2010 rankings, are absent in 2022. Four nations — Hong Kong, India, the Republic of Korea, and the United Arab Emirates — were not listed in 2010 and appeared in the 2022 ranking. Among these, Hong Kong and India are particularly prominent, securing the seventh and ninth positions, respectively.

In examining the Chemical section, the United States emerges as the most dominant entity, with China and Germany trailing closely behind. This analysis reveals a growing influence of the United States and China over the observed period, as indicated by the relative decline in the centrality metrics of other nations. Particularly noteworthy is the ascension through the ranks of the Netherlands, Belgium, Canada, Brazil, India, and the Republic of Korea, with India's rise to the fourth position by 2022 being especially remarkable. On the contrary, there is a noticeable descent in the standings of France, Great Britain, Japan, Italy, and Spain. Despite these shifts, the overall makeup of the Top 15 has shown remarkable consistency, except for Thailand's departure, which paved the way for Mexico's entry.

The centrality rankings for the Mineral section underscore China's ascendance as the preeminent nation within this network. Notably, the United States retained its second-place standing; however, its proximity to China's centrality value in 2010—amounting to 80\% of the leader's score—dwindled by 2022, falling to 45\%. This significant reduction accentuates the solidification of China's network influence. We draw attention to the upward trajectory of Australia, the Republic of Korea, France, and the Netherlands in the 2022 rankings. The Republic of Korea, in particular, merits acknowledgment for its climb from the eleventh to the seventh position, indicating a notable increase in its centrality within the network. Conversely, the ranking declined for Belgium, Hong Kong, Italy, Japan, and Singapore. Specifically, Hong Kong's descent from the eighth to the fourteenth position and Belgium's fall from the tenth to the fifteenth position are highlighted as significant shifts. Regarding ranking composition changes, we point out the exit of only two countries, Thailand and Canada, which were positioned towards the lower end of the Top 15 in 2010. The entry of India and Malaysia into the rankings is noteworthy, particularly India's impressive positioning at fourth place in 2022.

\begin{table*}[htbp]
\centering 
\begin{threeparttable}

\caption{Top 10 centrality rankings for the following sections: Mechanical \& Electrical, Chemical, and Mineral. The rankings are delineated in two separate panels, with the left panel showcasing the data for 2010 and the right panel for 2022.}
\label{tab:centrality1}
\centering
\begin{tabular}{cccccc}
\toprule\midrule
\multicolumn{3}{c}{$2010$} & \multicolumn{3}{c}{$2022$} \\
\cline{1-6}
Rank & \multicolumn{1}{l}{Centrality} & \multicolumn{1}{l}{Country} & \multicolumn{1}{l}{Rank} & \multicolumn{1}{l}{Centrality} & \multicolumn{1}{l}{Country} \\
\hline
\multicolumn{6}{c}{Mechanical \& Electrical} \\
\hline
1 &  \multicolumn{1}{l}{$100\%$} & \multicolumn{1}{l}{United States} & 1 & \multicolumn{1}{l}{$100\%$} & \multicolumn{1}{l}{United States} \\
2 &  \multicolumn{1}{l}{$57\%$} & \multicolumn{1}{l}{China} & 2 & \multicolumn{1}{l}{$46\%$} & \multicolumn{1}{l}{Germany} \\
3 & \multicolumn{1}{l}{$53\%$} & \multicolumn{1}{l}{Germany} & 3 & \multicolumn{1}{l}{$45\%$} & \multicolumn{1}{l}{China} \\
4 & \multicolumn{1}{l}{$50\%$} & \multicolumn{1}{l}{Thailand} & 4 & \multicolumn{1}{l}{$30\%$} & \multicolumn{1}{l}{Mexico} \\
5 & \multicolumn{1}{l}{$37\%$} & \multicolumn{1}{l}{Great Britain} & 5 & \multicolumn{1}{l}{$25\%$} & \multicolumn{1}{l}{France} \\
6 & \multicolumn{1}{l}{$33\%$} & \multicolumn{1}{l}{France} & 6 & \multicolumn{1}{l}{$25\%$} & \multicolumn{1}{l}{Canada} \\
7 & \multicolumn{1}{l}{$30\%$} & \multicolumn{1}{l}{Mexico} & 7 & \multicolumn{1}{l}{$24\%$} & \multicolumn{1}{l}{Hong Kong} \\
8 & \multicolumn{1}{l}{$28\%$} & \multicolumn{1}{l}{Netherlands} & 8 & \multicolumn{1}{l}{$22\%$} & \multicolumn{1}{l}{Netherlands} \\
9 & \multicolumn{1}{l}{$27\%$} & \multicolumn{1}{l}{Singapore} & 9 & \multicolumn{1}{l}{$21\%$} & \multicolumn{1}{l}{India} \\
10 & \multicolumn{1}{l}{$26\%$} & \multicolumn{1}{l}{Canada} & 10 & \multicolumn{1}{l}{$20\%$} & \multicolumn{1}{l}{Great Britain} \\
\hline
\multicolumn{6}{c}{Chemical} \\
\hline
1 &  \multicolumn{1}{l}{$100\%$} & \multicolumn{1}{l}{United States} & 1 & \multicolumn{1}{l}{$100\%$} & \multicolumn{1}{l}{United States} \\
2 &  \multicolumn{1}{l}{$64\%$} & \multicolumn{1}{l}{China} & 2 & \multicolumn{1}{l}{$65\%$} & \multicolumn{1}{l}{China} \\
3 & \multicolumn{1}{l}{$59\%$} & \multicolumn{1}{l}{Germany} & 3 & \multicolumn{1}{l}{$53\%$} & \multicolumn{1}{l}{Germany} \\
4 & \multicolumn{1}{l}{$43\%$} & \multicolumn{1}{l}{France} & 4 & \multicolumn{1}{l}{$31\%$} & \multicolumn{1}{l}{India} \\
5 & \multicolumn{1}{l}{$35\%$} & \multicolumn{1}{l}{Thailand} & 5 & \multicolumn{1}{l}{$30\%$} & \multicolumn{1}{l}{France} \\
6 & \multicolumn{1}{l}{$33\%$} & \multicolumn{1}{l}{Great Britain} & 6 & \multicolumn{1}{l}{$30\%$} & \multicolumn{1}{l}{Netherlands} \\
7 & \multicolumn{1}{l}{$32\%$} & \multicolumn{1}{l}{Japan} & 7 & \multicolumn{1}{l}{$27\%$} & \multicolumn{1}{l}{Brazil} \\
8 & \multicolumn{1}{l}{$32\%$} & \multicolumn{1}{l}{Netherlands} & 8 & \multicolumn{1}{l}{$25\%$} & \multicolumn{1}{l}{Belgium} \\
9 & \multicolumn{1}{l}{$31\%$} & \multicolumn{1}{l}{Belgium} & 9 & \multicolumn{1}{l}{$25\%$} & \multicolumn{1}{l}{Canada} \\
10 & \multicolumn{1}{l}{$27\%$} & \multicolumn{1}{l}{Canada} & 10 & \multicolumn{1}{l}{$24\%$} & \multicolumn{1}{l}{Great Britain} \\
\hline
\multicolumn{6}{c}{Mineral} \\
\hline
1 &  \multicolumn{1}{l}{$100\%$} & \multicolumn{1}{l}{China} & 1 & \multicolumn{1}{l}{$100\%$} & \multicolumn{1}{l}{China} \\
2 &  \multicolumn{1}{l}{$80\%$} & \multicolumn{1}{l}{United States} & 2 & \multicolumn{1}{l}{$45\%$} & \multicolumn{1}{l}{United States} \\
3 & \multicolumn{1}{l}{$53\%$} & \multicolumn{1}{l}{Singapore} & 3 & \multicolumn{1}{l}{$40\%$} & \multicolumn{1}{l}{Netherlands} \\
4 & \multicolumn{1}{l}{$50\%$} & \multicolumn{1}{l}{Japan} & 4 & \multicolumn{1}{l}{$37\%$} & \multicolumn{1}{l}{India} \\
5 & \multicolumn{1}{l}{$37\%$} & \multicolumn{1}{l}{Germany} & 5 & \multicolumn{1}{l}{$37\%$} & \multicolumn{1}{l}{Germany} \\
6 & \multicolumn{1}{l}{$35\%$} & \multicolumn{1}{l}{Netherlands} & 6 & \multicolumn{1}{l}{$34\%$} & \multicolumn{1}{l}{Singapore} \\
7 & \multicolumn{1}{l}{$29\%$} & \multicolumn{1}{l}{Italy} & 7 & \multicolumn{1}{l}{$34\%$} & \multicolumn{1}{l}{Republic of Korea} \\
8 & \multicolumn{1}{l}{$28\%$} & \multicolumn{1}{l}{Hong Kong} & 8 & \multicolumn{1}{l}{$33\%$} & \multicolumn{1}{l}{France} \\
9 & \multicolumn{1}{l}{$27\%$} & \multicolumn{1}{l}{France} & 9 & \multicolumn{1}{l}{$33\%$} & \multicolumn{1}{l}{Japan} \\
10 & \multicolumn{1}{l}{$22\%$} & \multicolumn{1}{l}{Belgium} & 10 & \multicolumn{1}{l}{$31\%$} & \multicolumn{1}{l}{Italy} \\
\midrule\toprule
\end{tabular}
    \begin{tablenotes}
    \footnotesize
      \item {Notes:} In our analysis, we have normalized the centrality values to facilitate a comparative understanding of each country's relative centrality within the specific section-level trade network. This normalization process adjusts the centrality scores, ensuring they are expressed in relation to the most central nation within each respective network.\\
    \end{tablenotes}
\end{threeparttable}

\end{table*}

Table~\ref{tab:centrality2} displays centrality rankings for the Transport and Base Metals sections. In Transport, Thailand's prominent role in 2010 diminished completely by 2022. The United States, ascending from second place in 2010, secured the leading position by 2022. This advancement positions the United States at the pinnacle of the section, followed closely by Germany and France, who climb to the second and third positions, respectively, thereby delineating the top echelon in 2022. Great Britain, Canada, and China improved their positioning in 2022, with China showing the most substantial climb from the eleventh to the sixth position, highlighting its growing influence in the Transport section. Conversely, Australia and Spain experienced a decline, with Australia dropping from the fifth to the fifteenth position. Composition shifts in the Top 15 of the Transport section is marked by the exit of five countries: Thailand, Russia, Saudi Arabia, Japan, Indonesia, and Brazil, introducing new participants such as Belgium, the United Arab Emirates, Poland, Mexico, and the Netherlands, thus reflecting the evolving landscape of this network.

In the Base Metals section, the United States maintains and amplifies its leadership, underlining a consolidation of influence as centrality values for the remainder of the Top 15 experience a noticeable decline in 2022. China and Germany remain steadfast, mirroring the top-tier dynamics observed in the Mechanical \& Electrical and Chemical sections. The trajectory for five countries—Italy, France, Japan, Great Britain, and the Netherlands—points downwards. However, Japan presents an anomaly, shifting from sixth place in 2010 to fifth in 2022, indicating a nuanced change in its network centrality. Conversely, Canada, Turkey, Mexico, and the Republic of Korea ascend within the rankings, with Mexico's commendable leap from twelfth to fourth position emphasizing its enhanced role in the network. The period under review witnesses the departure of Thailand and Belgium from the rankings, making room for India and Poland's entry. India, in particular, garners attention by securing the fifth position, showcasing its significant ascendancy within the Base Metals sector.

\begin{table*}[htbp]
\centering 
\begin{threeparttable}
\caption{In this table, we present the Top 15 centrality rankings for the following sections: Transport and Base Metals. The rankings are delineated in two separate panels, with the left panel showcasing the data for 2010 and the right panel for 2022.}
\label{tab:centrality2}
\centering
\begin{tabular}{cccccc}
\toprule\midrule
\multicolumn{3}{c}{$2010$} & \multicolumn{3}{c}{$2022$} \\
\cline{1-6}
Rank & \multicolumn{1}{l}{Centrality} & \multicolumn{1}{l}{Country} & \multicolumn{1}{l}{Rank} & \multicolumn{1}{l}{Centrality} & \multicolumn{1}{l}{Country} \\
\hline
\multicolumn{6}{c}{Transport} \\
\hline
1 &  \multicolumn{1}{l}{$100\%$} & \multicolumn{1}{l}{Thailand} & 1 & \multicolumn{1}{l}{$100\%$} & \multicolumn{1}{l}{United States} \\
2 &  \multicolumn{1}{l}{$48\%$} & \multicolumn{1}{l}{United States} & 2 & \multicolumn{1}{l}{$64\%$} & \multicolumn{1}{l}{Germany} \\
3 & \multicolumn{1}{l}{$46\%$} & \multicolumn{1}{l}{Germany} & 3 & \multicolumn{1}{l}{$39\%$} & \multicolumn{1}{l}{France} \\
4 & \multicolumn{1}{l}{$31\%$} & \multicolumn{1}{l}{France} & 4 & \multicolumn{1}{l}{$32\%$} & \multicolumn{1}{l}{Great Britain} \\
5 & \multicolumn{1}{l}{$25\%$} & \multicolumn{1}{l}{Australia} & 5 & \multicolumn{1}{l}{$32\%$} & \multicolumn{1}{l}{Canada} \\
6 & \multicolumn{1}{l}{$24\%$} & \multicolumn{1}{l}{Great Britain} & 6 & \multicolumn{1}{l}{$24\%$} & \multicolumn{1}{l}{China} \\
7 & \multicolumn{1}{l}{$17\%$} & \multicolumn{1}{l}{Canada} & 7 & \multicolumn{1}{l}{$24\%$} & \multicolumn{1}{l}{India} \\
8 & \multicolumn{1}{l}{$17\%$} & \multicolumn{1}{l}{Spain} & 8 & \multicolumn{1}{l}{$23\%$} & \multicolumn{1}{l}{Netherlands} \\
9 & \multicolumn{1}{l}{$16\%$} & \multicolumn{1}{l}{Italy} & 9 & \multicolumn{1}{l}{$20\%$} & \multicolumn{1}{l}{Italy} \\
10 & \multicolumn{1}{l}{$15\%$} & \multicolumn{1}{l}{Russia} & 10 & \multicolumn{1}{l}{$20\%$} & \multicolumn{1}{l}{Mexico} \\
11 & \multicolumn{1}{l}{$15\%$} & \multicolumn{1}{l}{China} & 11 & \multicolumn{1}{l}{$18\%$} & \multicolumn{1}{l}{Spain} \\
12 & \multicolumn{1}{l}{$14\%$} & \multicolumn{1}{l}{Saudi Arabia} & 12 & \multicolumn{1}{l}{$17\%$} & \multicolumn{1}{l}{Poland} \\
13 & \multicolumn{1}{l}{$13\%$} & \multicolumn{1}{l}{Japan} & 13 & \multicolumn{1}{l}{$17\%$} & \multicolumn{1}{l}{United Arabe Emirates} \\
14 & \multicolumn{1}{l}{$13\%$} & \multicolumn{1}{l}{Indonesia} & 14 & \multicolumn{1}{l}{$16\%$} & \multicolumn{1}{l}{Belgium} \\
15 & \multicolumn{1}{l}{$12\%$} & \multicolumn{1}{l}{Brazil} & 15 & \multicolumn{1}{l}{$15\%$} & \multicolumn{1}{l}{Australia} \\
\hline
\multicolumn{6}{c}{Base Metals} \\
\hline
1 &  \multicolumn{1}{l}{$100\%$} & \multicolumn{1}{l}{United States} & 1 & \multicolumn{1}{l}{$100\%$} & \multicolumn{1}{l}{United States} \\
2 &  \multicolumn{1}{l}{$85\%$} & \multicolumn{1}{l}{China} & 2 & \multicolumn{1}{l}{$57\%$} & \multicolumn{1}{l}{Germany} \\
3 & \multicolumn{1}{l}{$75\%$} & \multicolumn{1}{l}{Germany} & 3 & \multicolumn{1}{l}{$55\%$} & \multicolumn{1}{l}{China} \\
4 & \multicolumn{1}{l}{$42\%$} & \multicolumn{1}{l}{Italy} & 4 & \multicolumn{1}{l}{$34\%$} & \multicolumn{1}{l}{Mexico} \\
5 & \multicolumn{1}{l}{$39\%$} & \multicolumn{1}{l}{France} & 5 & \multicolumn{1}{l}{$32\%$} & \multicolumn{1}{l}{India} \\
6 & \multicolumn{1}{l}{$37\%$} & \multicolumn{1}{l}{Japan} & 6 & \multicolumn{1}{l}{$32\%$} & \multicolumn{1}{l}{Canada} \\
7 & \multicolumn{1}{l}{$37\%$} & \multicolumn{1}{l}{Canada} & 7 & \multicolumn{1}{l}{$31\%$} & \multicolumn{1}{l}{Italy} \\
8 & \multicolumn{1}{l}{$32\%$} & \multicolumn{1}{l}{Netherlands} & 8 & \multicolumn{1}{l}{$26\%$} & \multicolumn{1}{l}{France} \\
9 & \multicolumn{1}{l}{$32\%$} & \multicolumn{1}{l}{Thailand} & 9 & \multicolumn{1}{l}{$22\%$} & \multicolumn{1}{l}{Turkey} \\
10 & \multicolumn{1}{l}{$31\%$} & \multicolumn{1}{l}{Great Britain} & 10 & \multicolumn{1}{l}{$22\%$} & \multicolumn{1}{l}{Republic of Korea} \\
11 & \multicolumn{1}{l}{$29\%$} & \multicolumn{1}{l}{Turkey} & 11 & \multicolumn{1}{l}{$20\%$} & \multicolumn{1}{l}{Netherlands} \\
12 & \multicolumn{1}{l}{$28\%$} & \multicolumn{1}{l}{Mexico} & 12 & \multicolumn{1}{l}{$20\%$} & \multicolumn{1}{l}{Great Britain} \\
13 & \multicolumn{1}{l}{$28\%$} & \multicolumn{1}{l}{Republic of Korea} & 13 & \multicolumn{1}{l}{$18\%$} & \multicolumn{1}{l}{Poland} \\
14 & \multicolumn{1}{l}{$26\%$} & \multicolumn{1}{l}{Spain} & 14 & \multicolumn{1}{l}{$17\%$} & \multicolumn{1}{l}{Spain} \\
15 & \multicolumn{1}{l}{$23\%$} & \multicolumn{1}{l}{Belgium} & 15 & \multicolumn{1}{l}{$17\%$} & \multicolumn{1}{l}{Japan} \\
\midrule\toprule
\end{tabular}

    \begin{tablenotes}
    \footnotesize
      \item {Notes:} In our analysis, we have normalized the centrality values to facilitate a comparative understanding of each country's relative centrality within the specific section-level trade network. This normalization process adjusts the centrality scores, ensuring they are expressed in relation to the most central nation within each respective network.\\
    \end{tablenotes}

\end{threeparttable}

\end{table*}

\section{Trade network measures as predictors: a machine learning approach to GDP growth forecasting}

In this section, we explore the utilization of machine learning models for forecasting country-specific Gross Domestic Product (GDP) growth. Our analysis builds upon traditional methods that predict GDP growth using country-specific variables by introducing network descriptors extracted from section-specific trade networks. We believe network-based descriptors can offer predictive power for two main reasons. Firstly, a country's position in the trade network may convey information on its importance to other peers and its ability to access and substitute products and services across its counterparts. Secondly, economic conditions in a country's neighbors may provide insights into its own economic variables. For instance, if a neighbor transacts less than historically with a specific country, this behavior may signal potential GDP growth challenges for that country in the upcoming years. Both features can potentially affect a country's GDP growth. Therefore, we include network measures to introduce a novel dimension to the analysis, shedding light on the impact of the topological properties of international trade networks on economic forecasting.

The pursuit of accurate economic growth forecasting is an endeavor that profoundly influences strategic decision-making and policy formulation across the globe. It equips policymakers with the tools to design proactive measures to smooth out economic fluctuations, fostering a more stable economic environment. For businesses and investors, the insights gleaned from these forecasts anchor expectations and provide foresight that helps navigate market uncertainties and leverage opportunities for growth and expansion. Moreover, the importance of these forecasts extends beyond national borders. In an interconnected global economy, the ripple effects of economic trends in one country can have far-reaching impacts. Accurate forecasts enable countries to synchronize their economic policies and responses, essential for managing global challenges such as trade deficits and inflationary pressures.

We use a horse race to select among a set of supervised learning regression techniques. The suite of machine learning models applied includes the Light Gradient Boosting Machine (LightGBM), Support Vector Machines (SVM), k-nearest neighbor (k-NN) algorithms, Extreme Gradient Boosting (XGBoost), Linear Regression and its regularized variants, and Random Forest. This diverse array of models facilitates a thorough analysis by capturing the complex, both linear and non-linear, relationships between country-specific factors and the structure of section-level trade networks. 

In our model selection procedure, we use a cross-validation technique to search for each algorithm's most effective combination of hyperparameters, thereby optimizing predictive performance and ensuring model generalization to unseen examples. Hyperparameters, which are external configurations to the model that cannot be learned from the data, influence the training process and the architecture of the machine learning model. We undertake the hyperparameter tuning process by systematically navigating a predefined grid of hyperparameter values and evaluating model performance through adaptive resampling via futility analysis~\cite{kuhn2014futility}. We set the Root Mean Squared Error (RMSE) as our performance metric in the model selection. The model (and corresponding optimal set of hyperparameters) that minimizes the average RMSE across all models in the cross-validation procedure will be the winner in the horse race.


\subsection{Relevant background: supervised machine learning models}

This section reviews the suite of supervised regressions that constitute the horse race of algorithms.

\subsubsection{Linear Regression}

Linear Regression is a foundational statistical method employed for modeling the relationship between a dependent variable and one or more independent variables by optimizing the parameters of a linear equation best to fit the observed data~\cite{pearson1901liii,yule1897theory}. The core principle of Linear Regression lies in minimizing the sum of the squared differences between the observed outcomes in the dataset and the outcomes predicted by the linear approximation. Mathematically, this optimization problem can be expressed as minimizing the cost function \( J(\theta) = \frac{1}{2m} \sum_{i=1}^{m} \left(h_\theta(x^{(i)}) - y^{(i)}\right)^2 \), where $m$ denotes the number of observations, \( h_\theta(x)  \) represents the hypothesis function defined by \( h_\theta(x) = \theta^T x \), $y^{(i)}$ is the observed value, and 
$\theta$ are the parameters to be optimized. The algorithm selects $\theta$ such that it minimizes the cost function $J$ discussed above. Since $J$ is convex, the optimization process admits a closed-form solution, often called the ordinary least squares method. The linear regression has no hyperparameters.

\subsubsection{Elastic Net (Linear Regression with regularization)}

Regularized variants of Linear Regression, such as Ridge Regression (L2 regularization) and Lasso Regression (L1 regularization), introduce a penalty term to the cost function in the linear regression to prevent overfitting by constraining the magnitude of the model coefficients. Elastic Net Regression combines the strengths of both L1 and L2 regularization methods~\cite{zou2005regularization}. This approach is particularly effective in scenarios where multiple features are correlated and aims to balance the complexity of the model with its performance, thereby mitigating the issue of model overfitting. The cost function for Elastic Net Regression is formulated by incorporating both L1 and L2 penalty terms, optimizing the equation \(J(\theta) = \frac{1}{2m} \sum_{i=1}^{m} (h_\theta(x^{(i)}) - y^{(i)})^2 + \lambda_1 \sum_{j=1}^{n} |\theta_j| + \lambda_2 \sum_{j=1}^{n} \theta_j^2\), where \(\lambda_1\) and \(\lambda_2\) are the regularization parameters for the L1 and L2 penalties, respectively. This dual regularization approach allows Elastic Net to inherit the feature selection property of Lasso while also retaining the regularization benefits of Ridge, making it a versatile tool for regression analysis involving high-dimensional datasets. The hyperparameters we tune in the model selection are both $\lambda_1$ and $\lambda_2$.

\subsubsection{Support Vector Machine}

Support Vector Machine (SVM) with Radial Basis Function (RBF) kernel is a non-linear method~\cite{cortes1995support}. Unlike Linear Regression, which minimizes the discrepancies between observed and predicted values within a linear context, SVM with RBF kernel models complex, non-linear relationships in regression tasks. The SVM algorithm seeks to find the optimal hyperplane in a high-dimensional space that maximizes the margin (distance) of the hyperplane to the nearest points of each class, called support vectors. The RBF kernel is instrumental in transforming the input data into a higher-dimensional space, where linear regression becomes possible, thereby effectively handling non-linear data structures. The optimization function critical to SVM with RBF kernel for regression is formulated as minimizing \(J(\theta) = \frac{1}{2}||\theta||^2 + C \sum_{i=1}^{m} \left(\max(0, |y^{(i)} - (\theta^T \phi(x^{(i)}) + b)| - \epsilon)\right)^2\), where \(C\) is the regularization parameter, \(\phi(x^{(i)})\) represents the high-dimensional space mapped by the RBF kernel, \(b\) is the bias, and \(\epsilon\) denotes a margin of tolerance within which no penalty is given to errors. The parameter \(\gamma\) in the RBF kernel function, defined as \(\gamma = \frac{1}{2\sigma^2}\), plays a critical role in determining the decision boundary's flexibility. This optimization ensures that the model balances the complexity and the fitting accuracy, making SVM with RBF kernel a powerful tool for tackling non-linear regression challenges. The hyperparameters we tune in the model selection are \(C\) and \(\gamma\).

\subsubsection{k-Nearest Neighbor}

The k-Nearest Neighbor (k-NN) algorithm stands out in the machine learning landscape for its simplicity and non-parametric nature, contrasting sharply with the complexity of SVM with RBF kernel and the linearity of Linear Regression~\cite{cover1967nearest}. k-NN operates on the principle of feature similarity, predicting the outcome for a new instance based on the majority vote or average of its \(k\) closest neighbors in the feature space. This straightforward approach eliminates the need for parameter estimation, presenting an advantage in terms of simplicity and interpretability. Additionally, the performance heavily depends on the choice of \(k\) and the distance metric, which can significantly affect its accuracy. Unlike previous models, k-NN does not optimize a specific function for learning; instead, it directly uses the training data for prediction, minimizing an implicit cost function related to the distance between the query instance and its nearest neighbors, thereby determining the best fit for prediction. Despite its simplicity, k-NN's effectiveness is contingent upon a careful balance between the choice of \(k\) and the distance metric, ensuring adequate performance while highlighting its intuitive approach to machine learning prediction outcomes. The hyperparameter we tune in the model selection is \(k\).

\subsubsection{Random Forest}

The Random Forest algorithm is a prominent ensemble learning method designed to refine regression analyses through the collective predictions of multiple decision trees~\cite{breiman2001random}. This ensemble technique aims to elevate the robustness and accuracy of the model by merging the outputs from a variety of base learners, effectively mitigating the risk of overfitting common in more complex models. Specifically, Random Forest improves model generalization by averaging the individual predictions ($\hat{y}_n$) from each of the $N$ decision trees, making the ensemble's final prediction more reliable across diverse datasets. For a Random Forest composed of $N$ trees, the prediction $\hat{y}$ for any given input is calculated by averaging the outputs of all trees, represented as $\hat{y} = \frac{1}{N} \sum_{n=1}^{N} \hat{y}n$. The hyperparameters that we tune in the model selection are the number of trees ($N$), the maximum depth of each tree ($D$), and the number of features considered for splitting at each node ($F$).

\subsubsection{XGBoost}

XGBoost, an abbreviation for eXtreme Gradient Boosting, represents a sophisticated evolution of gradient boosting frameworks, acclaimed for its efficiency, versatility, and proficiency in processing large datasets~\cite{chen2016xgboost}. Unlike Random Forest, which generates independent trees in a parallel fashion, XGBoost constructs each tree sequentially. XGBoost distinguishes itself by adeptly managing linear and nonlinear datasets and incorporating regularization directly into its optimization process to mitigate overfitting. This regularization introduces penalties on the model's complexity, thereby balancing the reduction of prediction errors with the control of model complexity. The objective function optimized by XGBoost in regression tasks incorporates a regularized component, defined as \(J(\theta) = \sum_{i=1}^{m} l(y^{(i)}, \hat{y}^{(i)}) + \sum_{k} \Omega(f_k)\), where \(l\) signifies the loss function that measures the discrepancy between actual values \(y^{(i)}\) and predictions \(\hat{y}^{(i)}\), and \(\Omega\) denotes the regularization term affecting the complexity of the model's trees. The hyperparameters that we tune in the model selection are
the learning rate (\(\eta\)), the maximum depth of the trees (\(D\)) and the regularization parameters (\(\lambda\) for L2 regularization, \(\alpha\) for L1 regularization).

\subsubsection{Light Gradient Boosting Machine}

Light Gradient Boosting Machine (LightGBM), an advanced gradient boosting framework, excels in predictive modeling by optimizing a loss function $L = \sum_{i=1}^{n} l(y_i, \hat{y}_i)$, methodically incorporating weak learners in successive iterations through the formula $F_{k+1}(x) = F_k(x) + \alpha \cdot h(x)$~\cite{ke2017lightgbm}. It sets itself apart with unique features like Gradient-based One-Side Sampling (GOSS)~\footnote{GOSS (Gradient-based One-Side Sampling) gives preference to instances with larger gradients, indicating that they are more difficult to predict accurately, thereby focusing the learning process on these harder examples.} and Exclusive Feature Bundling (EFB)~\footnote{EFB (Exclusive Feature Bundling) aggregates sparse features that are mutually exclusive, reducing the dimensionality of the dataset with minimal loss of information.}, which substantially boost its computational efficiency. Directly comparing LightGBM to its counterpart, XGBoost, reveals LightGBM's advantage in terms of reduced memory usage and faster execution times thanks to these innovations while maintaining comparable levels of predictive accuracy. The approach to building trees sets LightGBM apart from XGBoost and other prevalent algorithms. Where many algorithms expand trees in a sequential, level-wise manner, LightGBM adopts a strategy focused on expanding trees by leaves, specifically targeting the leaf that is anticipated to result in the largest decrease in loss. In addition, while XGBoost and numerous other algorithms utilize a sorted-based method for learning decision trees—seeking the best division points among ordered attribute values—LightGBM employs a unique, histogram-based method. This model captures complex, non-linear relationships without requiring extensive feature engineering. This capability is fine-tuned through the adjustment of key hyperparameters such as the number of leaves ($L$), learning rate ($\eta$), and maximum depth of trees ($D$). These parameters play a crucial role in the model's performance, as encapsulated in the equation $F_{k+1}(x) = F_k(x) + \eta \cdot h_k(x, L, D)$, highlighting the sensitivity of LightGBM's output to these hyperparameters.

\subsection{Relevant background: the interpretable machine learning through Shapley values}

We evaluate the contributory significance of each feature within our predictive models by employing SHAP (SHapley Additive exPlanations) values~\cite{lundberg2017unified}. This tool is crucial for enhancing interpretability in machine learning models by extending the concept of Shapley values from cooperative game theory. Shapley values offer a systematic means to assess each feature's individual contribution towards a model's overall predictive accuracy. The allocation of contributions is determined through the examination of all possible permutations of features based on the Shapley value formula:

\begin{equation}
\phi_i(v) = \sum_{S \subseteq N \setminus \{i\}} \frac{|S|! (n - |S| - 1)!}{n!} \left( v(S \cup \{i\}) - v(S) \right)
\end{equation}

\noindent where $\phi_i(v)$ represents the SHAP value for feature $i$, $N$ is the set of all features, $n$ is the total number of features, $S$ is a subset of features excluding $i$, and $v(S)$ is the prediction function for subset $S$. This equation encapsulates the essence of SHAP values by quantifying the marginal contribution of feature $i$ when added to a subset of features $S$.

The principle of marginal contribution, pivotal to SHAP values, evaluates the additional impact of including a specific feature in a subset of features on the predictive outcome. SHAP values are calculated for each feature across all possible combinations, yielding a detailed portrayal of feature importance that is model-agnostic. This enables the interpretation of complex models by revealing the impact of each feature on the prediction for each instance in the dataset~\footnote{Some interesting application of SHAP values in Machine Learning models: mortality factors during COVID-19 pandemic~\cite{smith2021identifying}; estimation of Li-ion batteries~\cite{lee2022state} and prediction of bitcoin prices~\cite{giudici2021shapley}.}

\subsection{Pre-processing steps}

In our research, we execute several pre-processing steps on our dataset before applying machine learning models to ensure the integrity and quality of our analysis. Initially,  we transform all nominal predictor variables into factor variables.  Further refining our dataset, we filter out features exhibiting a correlation higher than 0.9 to mitigate the effects of multicollinearity, ensuring that highly correlated predictors do not skew our model's performance. Additionally, we eliminate features with near-zero variance, specifically those with a dominant category not sufficiently balanced by other categories, as identified by a frequency cut-off of 100. This step is essential for removing variables unlikely to contribute significantly to the model's predictive accuracy. 

We employ the k-NN imputation method to address missing values, leveraging the algorithm to estimate and replace missing data with the most plausible values based on similar observations. We also normalize all numeric predictors, excluding the "year" variable, to prevent variables with larger scales from overpowering those with smaller scales, ensuring each contributes equally to the model's efficacy. In the subsequent stage of our data preparation, we convert nominal predictors previously transformed into multiple dummy (binary) variables, each representing a category of the nominal predictor.  

\subsection{Results and discussion}

\subsubsection{Data description}

We utilize machine learning methodologies to forecast economic growth, designating annual GDP growth as our target variable. This data is acquired from the World Bank’s Open Data platform. We incorporate two distinct sets of feature variables into our analysis to achieve our objective of predicting economic growth. We note that all data employed in this analysis refers to annual values.

The initial set pertains to the fundamental characteristics of countries, which can further be categorized into relevant subsets. The sources of data for this analysis encompass the United Nations Conference on Trade and Development and the World Bank’s Open Data platform:

\begin{itemize}

\item Main economic indicators: GDP per capita (PPP); GDP (US\$); GDP deflator (\%); Gross capital formation to labor force; Unemployment rate (\%); Labour force participation rate (\%); Agriculture, forestry and fishing (\% of GDP); Manufacturing (\% of GDP); Services (\% of GDP); Carbon dioxide emissions per capita (tonnes); and Material footprint per capita (tonnes).

\vspace{5mm}

\item Economic openness: stimulates growth by fostering competition and efficiency, leading to technological advancements and productivity gains. It expands access to global markets, enhancing export opportunities and enabling firms to achieve economies of scale. Furthermore, it attracts foreign investment and facilitates knowledge transfer, contributing significantly to a country's economic development~\cite{huchet2018relationship}. We select the following variables: Trade (\% of GDP); Net trade in goods (US\$); Current account balance (\% of GDP); PPP conversion factor; Net inflows of foreign direct investment (\% of GDP); and Inward foreign direct investment stock.

\vspace{5mm}

\item Institutional quality: Governmental institutions lay the foundation for a country's economic activities. Effective institutions foster an environment conducive to economic engagement, innovation, growth, and development~\cite{butkiewicz2006institutional}. We employ measures of: Regulatory quality; Rule of law; Government effectiveness; Control of corruption; and Voice and Accountability.

\vspace{5mm}

\item Infrastructure: is a critical determinant of economic growth, acting as a catalyst for improving efficiency, productivity, and connectivity~\cite{esfahani2003institutions}. To account for the infrastructure quality, we employ: Fixed broadband subscriptions (per 100 people); Access to basic sanitation (\% population); and Urban population (\% population). 

\vspace{5mm}

\item Human development: Education significantly boosts economic growth by enhancing workforce productivity and fostering innovation, which leads to increased efficiency and groundbreaking advancements~\cite{benos2014education}. Health is a crucial driver of economic growth, primarily by enhancing labor productivity and reducing healthcare costs, thereby increasing economic efficiency and output~\cite{gyimah2004health}. We employ the following measures:  Mean years of Schooling; Human development index; Human inequality; Life expectancy at birth; Refugee population; and Population growth (\%).

\vspace{5mm}

\end{itemize}

The second set of feature variables we introduce comprises network measures extracted from section-level international trade networks outlined in Section~\ref{sec:net_anal}.  These metrics are considered across various sections of trade, namely Mechanical \& Electrical, Mineral, Transport, Chemical, and Base Metals. For an in-depth exploration of the data utilized to construct these networks and the derivation of their metrics, readers are directed to Subsection~\ref{subssec_netmeas}.

\subsubsection{Predictive accuracy}

We run the horse race of supervised algorithms discussed before. We use the RMSE as the criterion to select the best model. However, we also report other metrics to show that the selected algorithm is robust when we compare other performance metrics: Huber Loss, Mean Absolute Error (MAE), Root Mean Square Error (RMSE), and Symmetric Mean Absolute Percentage Error (SMAPE)~\footnote{The Huber Loss merges squared loss for small deviations and absolute loss for larger ones, effectively managing outliers with less severe penalties. This trait is beneficial for economic data prone to outliers from economic shocks or errors. MAE calculates the average error magnitude, offering straightforward interpretability without weighing the direction of errors. Its limitation is the equal treatment of all errors, potentially overlooking significant prediction mistakes. RMSE amplifies larger errors by squaring them, enhancing sensitivity to outliers, and spotlighting models with occasional inaccurate predictions, but this also means RMSE is less tolerant of outliers than MAE. SMAPE uses percentage errors, ensuring scale independence and suitability for comparing different datasets, although its reliability falters with values near zero, restricting its use in certain cases.}.  Achieving lower values across these metrics signifies superior predictive accuracy, which indicates that the model's forecasts are consistently closer to the actual observed values.

\begin{figure*}[!ht]
  \centering
    \includegraphics[width=0.95\textwidth]{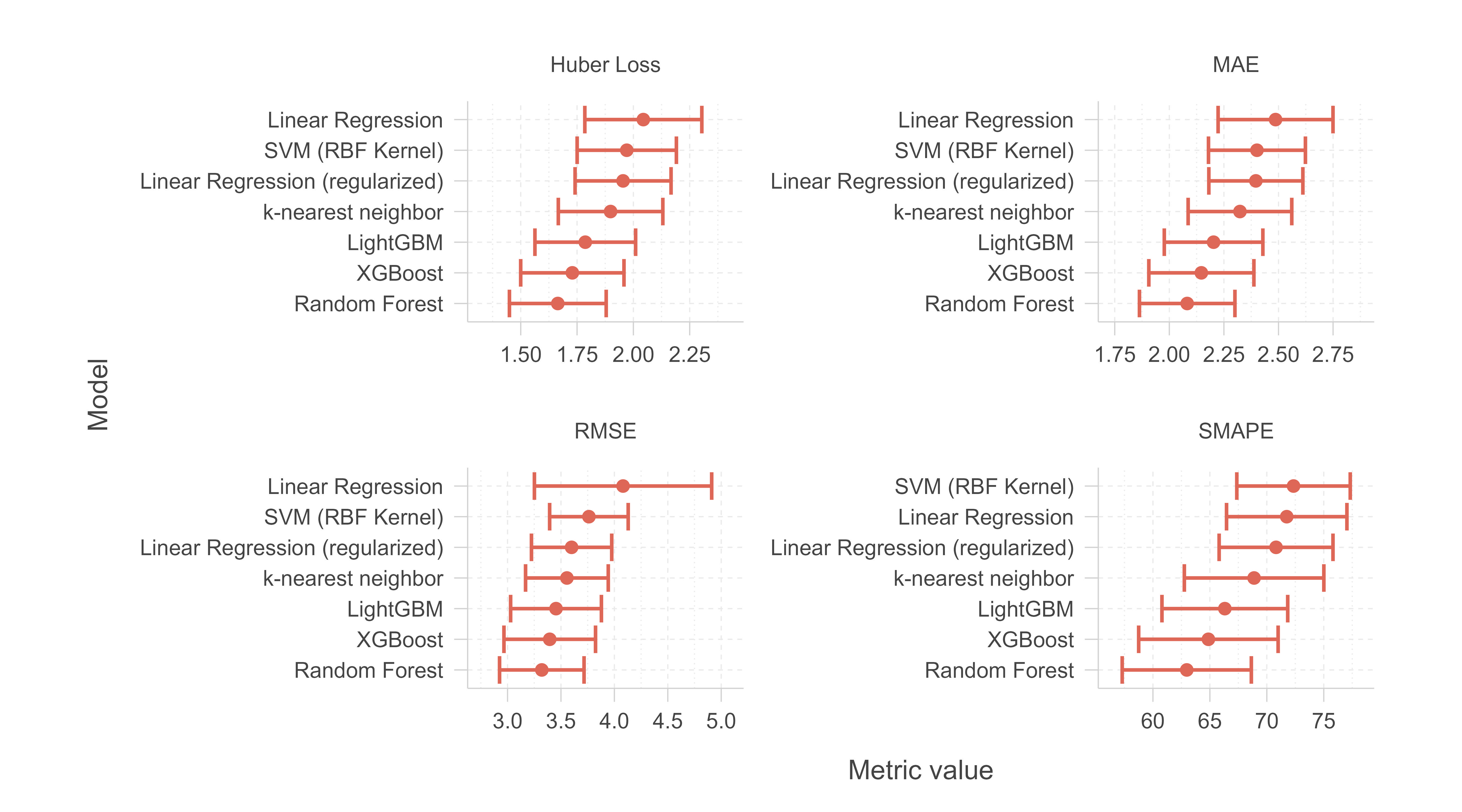}
  \caption{This figure illustrates the estimated error values and their associated confidence intervals for the four error metrics—Huber Loss, Mean Absolute Error (MAE), Root Mean Square Error (RMSE), and Symmetric Mean Absolute Percentage Error (SMAPE)—evaluated across the seven machine learning algorithms examined in our study.}
  \label{fig:errors}
\end{figure*}

In Figure~\ref{fig:errors}, we detail the average error values and their corresponding 95\% confidence intervals across the four error metrics, evaluated for the seven machine learning models included in our study over each of the out-of-sample remaining fold in the cros-validation procedure.  The analysis elucidates a clear hierarchical structure in model performance, with the Random Forest model leading and closely followed by XGBoost and LightGBM~\footnote{The optimal hyperparameter configurations for our suite of machine learning models are as follows: For the Elastic Net model, the optimal penalty is 0.007739036873 and the mixture parameter is 0.6662955002. For the SVM (RBF kernel), the cost is 0.5971530254 and the RBF sigma is 0.003403106644. The k-NN model requires 11 neighbors, utilizing an inverse distance weighting function and a distance power of 0.9160159697. The Random Forest model has an mtry value of 39—indicating the number of features considered for splitting at each tree node—alongside a minimum node size of 12. For XGBoost, there are 861 trees and a learning rate of 0.01860805586, whereas the LightGBM model requires 1107 trees and a learning rate of 0.03441793697.}. Subsequent positions are occupied by k-NN, regularized Linear Regression, SVM with RBF Kernel, and, finally, Linear Regression. Generally, this ranking is maintained across the performance metrics. In our next section, we attempt to interpret the model's predictions using the best-performing models in the horse race. 


\subsubsection{Feature importance}

\begin{figure}[htbp]
  \centering
    \includegraphics[width=0.6\textwidth]{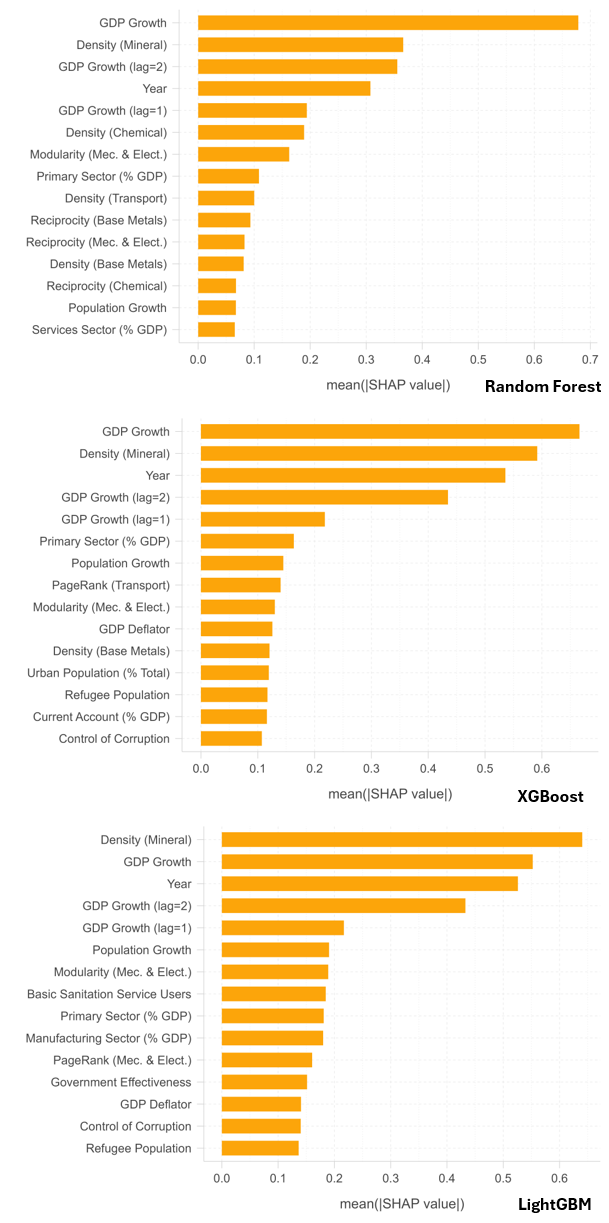}
  \caption{This figure illustrates the average SHAP values for the Top 15 most influential features as determined for the XGBoost, LightGBM, and Random Forest models. The visualization provides a comparative analysis across the three models, highlighting how each model values different features in terms of their predictive power. }
  \label{fig:model_shap}
\end{figure}

Figure~\ref{fig:model_shap} displays the average SHAP values in the module for the Top 15 most influential features as identified across the three highest-performing models in our analysis. Notably, current GDP growth emerges as the paramount feature in both the Random Forest and XGBoost models, whereas it ranks second in the LightGBM model. A particularly striking observation is the pronounced dominance of this feature in the Random Forest model, where its SHAP value is approximately twice as large as that of the second-ranking feature. The density of the Mineral trade network is the second most critical feature for the Random Forest and XGBoost models, surpassing current GDP growth in the LightGBM model. Temporal dynamics rank as the third most significant feature for both the XGBoost and LightGBM models and fourth for the Random Forest model. The two-period lagged GDP growth rate secures the fourth position in the XGBoost and LightGBM models, while it climbs to the third spot in the Random Forest model.

This initial analysis yields several insightful observations. Primarily, these four variables collectively form a core group that exhibits considerably large SHAP values relative to other features, highlighting their pivotal role in forecasting economic growth. Secondly, the significant positioning of both current and two-period lagged GDP growth rates suggests an "economic inertia" effect, indicating that present and recent economic performances have implications for future growth trajectories. Thirdly, the connectivity of trade networks plays a critical role in economic growth predictions, with particular emphasis on the Mineral section.

A notable discovery is the substantial representation of network metrics among the Top 15 features within the Random Forest model, the most effective model of our work. Remarkably, more than half of these leading features pertain to network metrics, with eight out of fifteen. Specifically, the metrics of density and reciprocity are especially prominent, constituting seven out of these eight network metrics. Furthermore, we observe that each of the five commodity sections covered in our study features at least one network metric among the top-ranking features, suggesting their overall significance in economic growth forecasts. This prevalence starkly contrasts with the findings from the other two models, where network metrics are considerably less represented, making up only three out of the Top 15 features.

Moreover, our research consistently highlights the importance of three variables across all three models: the modularity of the Mechanical \& Electrical trade network, population growth, and the value added by agriculture, forestry, and fishing to GDP. This consistency across varied models underscores the significance of network connectivity, population dynamics, and the contributions of primary sector activities to predicting economic growth.

\subsubsection{In-depth analysis of the Random Forest model}

Our analysis shows that the Random Forest model is superior in predictive accuracy. Consequently, we select this model for an in-depth examination with regard to feature interpretability using Shapley values. We develop a beeswarm dependence plot to illustrate the Top 20 features, identified by the highest mean absolute SHAP values. The vertical axis of the plot orders features by their significance in the model, while the horizontal axis displays standardized SHAP values. Each dot in the plot symbolizes the marginal contribution of a specific attribute for an individual observation, enabling the examination of SHAP value distributions through denser (indicating a higher concentration of observations) and sparser areas (indicating a lower concentration). Positive SHAP values indicate a positive contribution (increases) to the GDP growth in the following year, whereas negative values indicate a negative contribution (decreases) of the feature to the GDP growth in the following year. The coloration of dots indicates the attribute's value: lighter (darker) colors represent higher (lower) values for the attribute.


Our scrutiny begins with the most relevant feature: the current GDP growth. Observations with lower values of this feature have negative SHAP values, while higher values align with positive ones. This suggests that reduced current GDP growth rates portend a lower GDP growth next year and vice versa. This pattern is consistent with examining the two-period lagged GDP growth, reinforcing the "economic inertia" concept. This concurs with economic literature, which posits that economic growth exhibits an "autoregressive component". This rationalizes the widespread use in the literature of autoregressive models as a baseline for comparison with more complex models~\cite{camba2001automatic}.

The density of the Mineral trade network is the second most relevant attribute. The variability in this feature's values is relatively low, as indicated by the dot coloration. Medium values are associated with positive SHAP values, whereas higher values correspond to negative SHAP values, suggesting an intriguing pattern: moderate to high network density values correlate with positive future economic growth, yet exceedingly high values forecast negative economic outcomes, implying a threshold beyond which network density inversely affects economic growth in the following year. This link between the Mineral trade network's metric and future economic performance may be attributed to the influence of commodity prices—such as iron ore, oil, and coal—on business cycles~\cite{crucini2011driving}. Similar patterns are observed for the density measures of Chemical, Transport, and Base Metals trade networks.

A significant discovery relates to the feature associated with the primary sector. High values of this feature have positive SHAP values, while lower values align with negative SHAP values, indicating that an enhanced relevance of the primary sector within an economy forecasts positive economic growth in the following year and vice versa. This aligns with literature stating the crucial role of agriculture in shaping aggregate business cycles across countries~\cite{da2006role}~\footnote{Economies heavily reliant on agriculture experience pronounced fluctuations in overall output, exhibit low relative volatility in aggregate employment, and demonstrate a weak correlation between output and employment. Moreover, agriculture is distinctively characterized by more volatile output and employment, which are not positively correlated with those in other sectors, and a weaker correlation between output and employment within agriculture than in non-agricultural sectors. These unique characteristics significantly influence the dynamics of aggregate business cycles across countries~\cite{da2006role}.}. Additionally, the patterns associated with population growth are noteworthy. Higher population growth values align with positive Shapley values, whereas lower values predict negative economic growth in the following year. This outcome aligns intuitively with conventional models of economic growth that posit the relevant role of population dynamics in driving economic performance~\cite{becker1999population}.

\begin{figure}[htbp]
  \centering
    \includegraphics[width=0.98\textwidth]{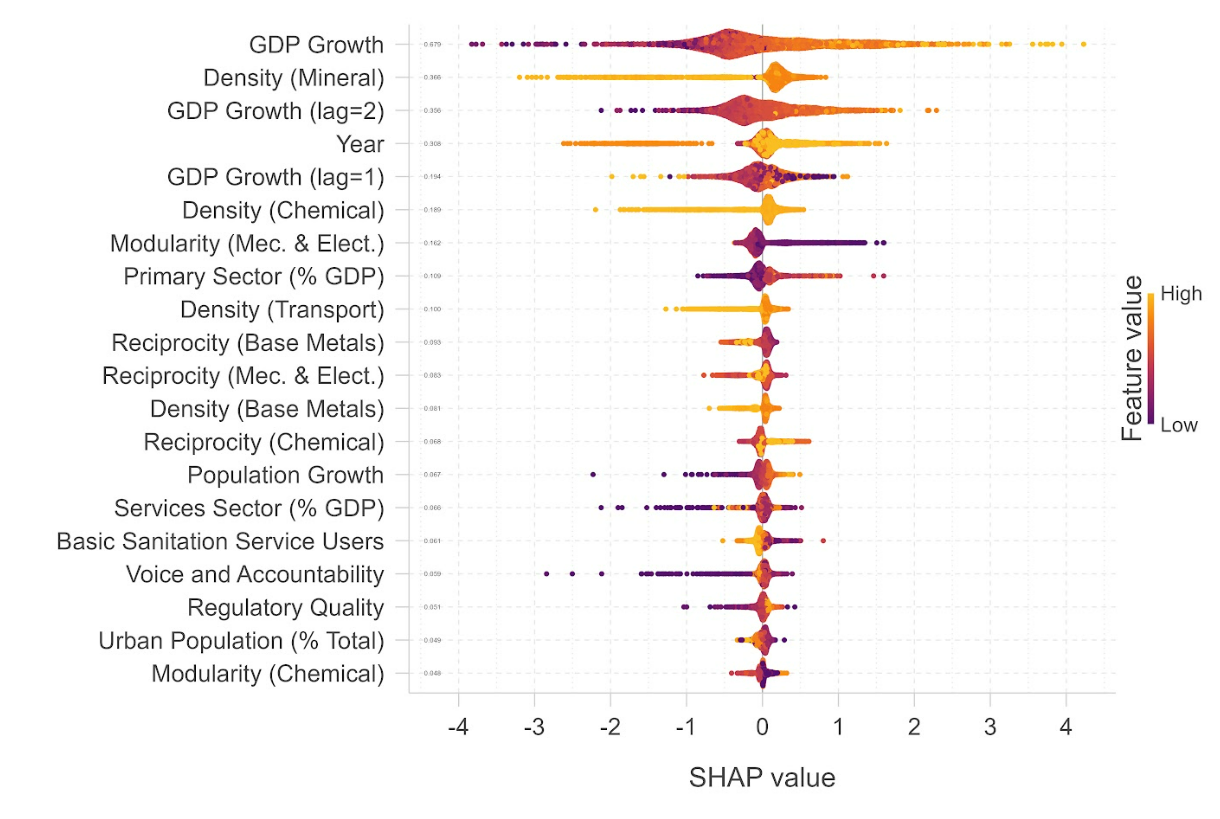}
  \caption{This figure displays a beeswarm plot of the Top 20 features from the Random Forest model, ranked by their mean SHAP values. Features are ordered vertically by importance, with their standardized SHAP values plotted horizontally, showing the contribution to prediction. Dots represent individual observations, with their distribution indicating the variability of SHAP values; dense areas suggest higher concentration, while sparse areas indicate less. The color of the dots reflects the feature values, elucidating the influence of high versus low feature values on economic growth forecasts, with positive SHAP values signaling potential growth and negative values indicating possible downturns.}
  \label{fig:beworm}
\end{figure}

\begin{figure}[htbp]
  \centering
    \includegraphics[width=0.63\textwidth]{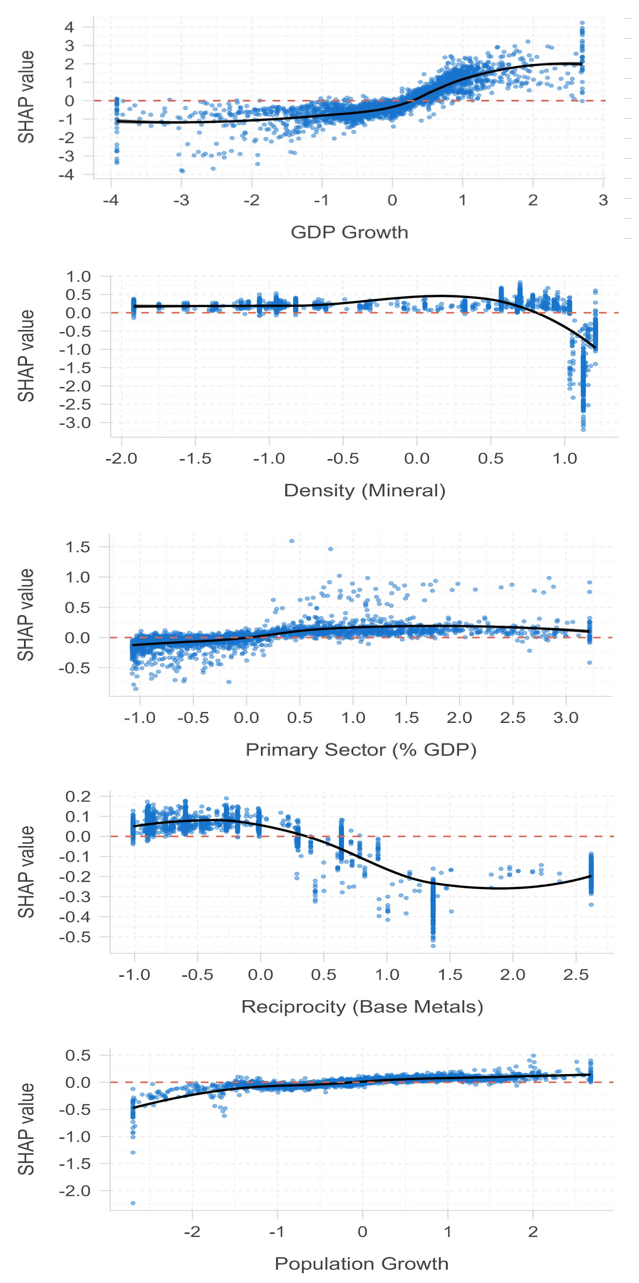}
  \caption{This figure showcases the SHAP value dependence plots for five key features of the Random Forest model. These plots elucidate the nuanced relationships between variations in feature values and their impact on predictive outcomes. By standardizing feature values along the horizontal axis, the plots effectively demonstrate how both positive and negative deviations from the mean feature value contribute to changes in the model's economic growth predictions.}
  \label{fig:dep_plot}
\end{figure}

We conclude our analysis by presenting in Figure~\ref{fig:dep_plot} the SHAP value dependence plots for five selected features of the Random Forest model. These plots augment the insights gained from the beeswarm dependence plot analysis by offering a nuanced view of how variations in feature values influence economic growth forecasts. On the horizontal axis, feature values are standardized, providing a vivid visual representation of how deviations from the average feature value—both positive and negative—affect predictions.

Commencing with the plot for current GDP growth, we observe that an increase of 1 standard deviation from the average current GDP growth induces a significant uptick in SHAP value. Further increments of a similar magnitude yield only marginally higher SHAP value, indicating a trend of diminishing marginal contributions to GDP growth in the following year. In contrast, a decrease of 1 standard deviation from the mean value leads to a decline in SHAP value, venturing into negative territory. Notably, SHAP values demonstrate lesser sensitivity to decreases of 1 standard deviation than to increases of the same magnitude relative to the mean value. Moreover, our analysis reveals a predominantly linear decline in SHAP values as decreases extend to 3 standard deviations below the mean. 

Shifting the focus to the network density of the Mineral trade network, we identify patterns distinctly different from those associated with current GDP growth. Variations in network density ranging from -2 to about -0.75 standard deviations from the average feature value exert no influence on SHAP values. However, within the span of -0.75 to approximately 0.25 standard deviations from the mean, SHAP values trend positively. Elevating network density beyond 0.4 standard deviations from the mean triggers a marked negative trajectory, with SHAP values descending into increasingly negative realms. This pattern underscores a complex relationship: while moderate increases in network density within the mineral section are correlated with neutral or positive effects on economic growth in the following year, surpassing a critical threshold links further increases in network density with declining future economic prospects, suggesting an optimal range for network density enhancements to positively influence future economic growth.

Subsequently, we explore an additional network metric: reciprocity of the Base Metals section. Our analysis delineates three distinct intervals, each characterized by varying behaviors of SHAP values in response to changes in reciprocity. In the interval between -1 and -0.25 standard deviations, the SHAP values manifest a positive trend, where increments in reciprocity correlate with higher SHAP values. Conversely, in the range from -0.25 to 2 standard deviations, an increase in reciprocity decreases SHAP values, which ultimately become negative. Finally, within the interval spanning from -2 to -2.5 standard deviations, although increases in reciprocity are associated with elevated SHAP values, these adjustments are insufficient to revert them to positive figures. This observation underscores a notable result: reciprocity values below the average are associated with positive GDP growth in the subsequent year, whereas reciprocity levels above the average indicate negative GDP growth forecasts.

In our evaluation of the value added by agriculture, forestry, and fishing activities to GDP, positive deviations within the range of 0 to 1 standard deviation from the mean yield positive and ascending SHAP values. Between 1 to 2 standard deviations, SHAP values level off, signifying a stabilization in their predictive impact. Beyond two standard deviations, we observe a gradual descent in SHAP values. Conversely, intervals between -1 and 0 standard deviations from the mean feature value show negative SHAP values, following an approximately linear decline. These observations suggest that positive deviations from the mean can enhance economic growth forecasts (up to two standard deviations), while negative deviations are associated with diminished economic prospects.

Lastly, we analyze population growth, identifying a linear relationship in SHAP values within the range of -1.5 to 2 standard deviations from the mean feature value. Values below the mean correlate with negative SHAP values, and those above the mean with positive SHAP values. We note an accelerated decline in SHAP value for feature values descending beyond -1.5 standard deviations from the mean. This indicates that within the predominant range of feature values, the relationship between population growth and economic growth in the following year is approximately linear, except at significant negative deviations from the mean, where the impact on economic growth forecasts intensifies.

\section{Conclusions}

We reflect on our efforts to elucidate the intricate relationship between international trade networks and machine learning techniques in forecasting economic growth. Our objective was to transcend traditional forecasting methodologies by harnessing the predictive power of network topological measures alongside advanced machine learning models. Central to our findings is recognizing the necessity for models that can unravel complex patterns and non-linear relationships while integrating international trade networks as pivotal features to refine economic growth forecasts.

Our detailed analysis began with constructing section-level international trade networks and extracting topological measures, which unveiled significant transformations within these networks amidst de-globalization trends. The pivotal period identified between 2016 and 2018, marked by a reversal of prior trends, highlighted the deep-seated impact of trade policy uncertainties intensified by global events such as the USA-China trade war, the COVID-19 pandemic, and Russia's invasion of Ukraine. The centrality rankings we developed exposed shifting paradigms of influence within the international trade domain, emphasizing the sustained dominance of certain nations and the notable rise of others, like India, in key commodity-specific trade networks.

Moreover, our venture into machine learning for economic growth forecasting revealed the superior performance of models such as Random Forest, XGBoost, and k-NN over traditional linear regression models. This finding confirms the complex and non-linear patterns these advanced models capture and highlights network measures' critical role in boosting forecasts' accuracy. In particular, the density of the Mineral trade network stood out as a pivotal predictor, showcasing the nuanced interplay between trade network dynamics and economic forecasting.

Our feature importance analysis highlighted the autoregressive nature of economic growth, underscoring the imperative of grasping both past and present GDP dynamics to predict future economic performance effectively. This insight is enriched further by identifying key growth predictors, such as the modularity of the Machine \& Electrical trade network, population growth, and the significance of the primary sector. The consistent significance of these factors across the top-performing models underscores their essential role in predicting future economic performance.

Delving into feature interpretability, particularly through the Random Forest model, we uncovered the complex ways pivotal features influence economic forecasts. The use of SHAP value dependence plots proved crucial in deciphering these patterns, for example, demonstrating how variations in the Mineral trade network's density correlate with GDP growth predictions. Identifying an optimal network density range highlights the need for nuanced policy and decision-making to leverage network topological analysis benefits while mitigating potential drawbacks.

Looking to the future, there is ample scope for expanding this research. Delving deeper into the causal links between network topologies and economic outcomes, incorporating real-time data, and exploring predictive models in light of sudden economic shocks or geopolitical developments could offer critical insights for policymakers striving for resilience and adaptability in an ever-changing and dynamic global landscape.

 


\bibliographystyle{ieeetr}
\bibliographystyle{abbrv}

\begin{thebibliography}{10}

\bibitem{anghinoni2019time}
L.~Anghinoni, L.~Zhao, D.~Ji, and H.~Pan.
\newblock Time series trend detection and forecasting using complex network topology analysis.
\newblock {\em Neural Networks}, 117:295--306, 2019.

\bibitem{barthelemy2005characterization}
M.~Barth{\'e}lemy, A.~Barrat, R.~Pastor-Satorras, and A.~Vespignani.
\newblock Characterization and modeling of weighted networks.
\newblock {\em Physica a: Statistical mechanics and its applications}, 346(1-2):34--43, 2005.

\bibitem{becker1999population}
G.~S. Becker, E.~L. Glaeser, and K.~M. Murphy.
\newblock Population and economic growth.
\newblock {\em American Economic Review}, 89(2):145--149, 1999.

\bibitem{benos2014education}
N.~Benos and S.~Zotou.
\newblock Education and economic growth: A meta-regression analysis.
\newblock {\em World Development}, 64:669--689, 2014.

\bibitem{breiman2001random}
L.~Breiman.
\newblock Random forests.
\newblock {\em Machine Learning}, 45:5--32, 2001.

\bibitem{butkiewicz2006institutional}
J.~L. Butkiewicz and H.~Yanikkaya.
\newblock Institutional quality and economic growth: Maintenance of the rule of law or democratic institutions, or both?
\newblock {\em Economic Modelling}, 23(4):648--661, 2006.

\bibitem{caldara2020economic}
D.~Caldara, M.~Iacoviello, P.~Molligo, A.~Prestipino, and A.~Raffo.
\newblock The economic effects of trade policy uncertainty.
\newblock {\em Journal of Monetary Economics}, 109:38--59, 2020.

\bibitem{camba2001automatic}
G.~Camba-Mendez, G.~Kapetanios, R.~J. Smith, and M.~R. Weale.
\newblock An automatic leading indicator of economic activity: forecasting gdp growth for european countries.
\newblock {\em The Econometrics Journal}, 4(1):S56--S90, 2001.

\bibitem{chen2016xgboost}
T.~Chen and C.~Guestrin.
\newblock Xgboost: A scalable tree boosting system.
\newblock In {\em 22nd ACM SIGKDD International Conference on Knowledge Discovery and Data Mining}, pages 785--794, 2016.

\bibitem{chi2010network}
K.~T. Chi, J.~Liu, and F.~C. Lau.
\newblock A network perspective of the stock market.
\newblock {\em Journal of Empirical Finance}, 17(4):659--667, 2010.

\bibitem{chuku2019intelligent}
C.~Chuku, A.~Simpasa, and J.~Oduor.
\newblock Intelligent forecasting of economic growth for developing economies.
\newblock {\em International Economics}, 159:74--93, 2019.

\bibitem{chuluun2017firm}
T.~Chuluun, A.~Prevost, and A.~Upadhyay.
\newblock Firm network structure and innovation.
\newblock {\em Journal of Corporate Finance}, 44:193--214, 2017.

\bibitem{cortes1995support}
C.~Cortes and V.~Vapnik.
\newblock Support-vector networks.
\newblock {\em Machine Learning}, 20:273--297, 1995.

\bibitem{cover1967nearest}
T.~Cover and P.~Hart.
\newblock Nearest neighbor pattern classification.
\newblock {\em IEEE Transactions on Information Theory}, 13(1):21--27, 1967.

\bibitem{crucini2011driving}
M.~J. Crucini, M.~A. Kose, and C.~Otrok.
\newblock What are the driving forces of international business cycles?
\newblock {\em Review of Economic Dynamics}, 14(1):156--175, 2011.

\bibitem{da2006role}
J.~M. Da-Rocha and D.~Restuccia.
\newblock The role of agriculture in aggregate business cycles.
\newblock {\em Review of Economic Dynamics}, 9(3):455--482, 2006.

\bibitem{de200625}
J.~G. De~Gooijer and R.~J. Hyndman.
\newblock 25 years of time series forecasting.
\newblock {\em International Journal of Forecasting}, 22(3):443--473, 2006.

\bibitem{demirer2018estimating}
M.~Demirer, F.~X. Diebold, L.~Liu, and K.~Yilmaz.
\newblock Estimating global bank network connectedness.
\newblock {\em Journal of Applied Econometrics}, 33(1):1--15, 2018.

\bibitem{elliott2008economic}
G.~Elliott and A.~Timmermann.
\newblock Economic forecasting.
\newblock {\em Journal of Economic Literature}, 46(1):3--56, 2008.

\bibitem{esfahani2003institutions}
H.~S. Esfahani and M.~T. Ramirez.
\newblock Institutions, infrastructure, and economic growth.
\newblock {\em Journal of Development Economics}, 70(2):443--477, 2003.

\bibitem{gao2015features}
C.~Gao, M.~Sun, and B.~Shen.
\newblock Features and evolution of international fossil energy trade relationships: A weighted multilayer network analysis.
\newblock {\em Applied Energy}, 156:542--554, 2015.

\bibitem{garlaschelli2004fitness}
D.~Garlaschelli and M.~I. Loffredo.
\newblock Fitness-dependent topological properties of the world trade web.
\newblock {\em Physical Review Letters}, 93(18):188701, 2004.

\bibitem{giudici2021shapley}
P.~Giudici and E.~Raffinetti.
\newblock Shapley-lorenz explainable artificial intelligence.
\newblock {\em Expert systems with applications}, 167:114104, 2021.

\bibitem{goulet2022machine}
P.~Goulet~Coulombe, M.~Leroux, D.~Stevanovic, and S.~Surprenant.
\newblock How is machine learning useful for macroeconomic forecasting?
\newblock {\em Journal of Applied Econometrics}, 37(5):920--964, 2022.

\bibitem{gyimah2004health}
K.~Gyimah-Brempong and M.~Wilson.
\newblock Health human capital and economic growth in sub-saharan african and oecd countries.
\newblock {\em The quarterly review of economics and finance}, 44(2):296--320, 2004.

\bibitem{hegre2010trade}
H.~Hegre, J.~R. Oneal, and B.~Russett.
\newblock Trade does promote peace: New simultaneous estimates of the reciprocal effects of trade and conflict.
\newblock {\em Journal of Peace Research}, 47(6):763--774, 2010.

\bibitem{hu2020characteristics}
X.~Hu, C.~Wang, M.~K. Lim, and W.-Q. Chen.
\newblock Characteristics of the global copper raw materials and scrap trade systems and the policy impacts of china's import ban.
\newblock {\em Ecological economics}, 172:106626, 2020.

\bibitem{huchet2018relationship}
M.~Huchet-Bourdon, C.~Le~Mou{\"e}l, and M.~Vijil.
\newblock The relationship between trade openness and economic growth: Some new insights on the openness measurement issue.
\newblock {\em The World Economy}, 41(1):59--76, 2018.

\bibitem{ke2017lightgbm}
G.~Ke, Q.~Meng, T.~Finley, T.~Wang, W.~Chen, W.~Ma, Q.~Ye, and T.-Y. Liu.
\newblock Lightgbm: A highly efficient gradient boosting decision tree.
\newblock {\em Advances in Neural Information Processing Systems}, 30, 2017.

\bibitem{kuhn2014futility}
M.~Kuhn.
\newblock Futility analysis in the cross-validation of machine learning models.
\newblock {\em arXiv preprint arXiv:1405.6974}, 2014.

\bibitem{lee2022state}
G.~Lee, J.~Kim, and C.~Lee.
\newblock State-of-health estimation of li-ion batteries in the early phases of qualification tests: An interpretable machine learning approach.
\newblock {\em Expert Systems with Applications}, 197:116817, 2022.

\bibitem{lee2019global}
T.~K. Lee, J.~H. Cho, D.~S. Kwon, and S.~Y. Sohn.
\newblock Global stock market investment strategies based on financial network indicators using machine learning techniques.
\newblock {\em Expert Systems with Applications}, 117:228--242, 2019.

\bibitem{lovric2018social}
M.~Lovri{\'c}, R.~Da~Re, E.~Vidale, D.~Pettenella, and R.~Mavsar.
\newblock Social network analysis as a tool for the analysis of international trade of wood and non-wood forest products.
\newblock {\em Forest Policy and Economics}, 86:45--66, 2018.

\bibitem{lundberg2017unified}
S.~M. Lundberg and S.-I. Lee.
\newblock A unified approach to interpreting model predictions.
\newblock {\em Advances in Neural Information Processing Systems}, 30, 2017.

\bibitem{ma2022analysis}
Y.~Ma, M.~Wang, and X.~Li.
\newblock Analysis of the characteristics and stability of the global complex nickel ore trade network.
\newblock {\em Resources Policy}, 79:103089, 2022.

\bibitem{masini2023machine}
R.~P. Masini, M.~C. Medeiros, and E.~F. Mendes.
\newblock Machine learning advances for time series forecasting.
\newblock {\em Journal of Economic Surveys}, 37(1):76--111, 2023.

\bibitem{Note1}
This body of work is extensive and includes analyses such as the behavior of stock returns amidst the Global Financial Crisis~\cite {silva2023effect}, the fluctuation patterns in stock prices~\cite {chi2010network}, and the intricate relationship between the advent of COVID-19 and stock market dynamics~\cite {samitas2022covid}. Further exploration within this field includes the examination of the financial impacts stemming from shifts in monetary policy~\cite {silva2020micro} and the formulation of investment strategies within a global framework~\cite {lee2019global}. The literature also navigates through the interconnectedness of global banking networks~\cite {demirer2018estimating}, the correlation between a firm's position in the supply chain structure and its innovation capabilities~\cite {chuluun2017firm}, and the evaluation of financial system vulnerabilities and the spread of economic impacts~\cite {silva2017monitoring}.

\bibitem{Note10}
EFB (Exclusive Feature Bundling) aggregates sparse features that are mutually exclusive, reducing the dimensionality of the dataset with minimal loss of information.

\bibitem{Note11}
Some interesting application of SHAP values in Machine Learning models: mortality factors during COVID-19 pandemic~\cite {smith2021identifying}; estimation of Li-ion batteries~\cite {lee2022state} and prediction of bitcoin prices~\cite {giudici2021shapley}.

\bibitem{Note12}
The Huber Loss merges squared loss for small deviations and absolute loss for larger ones, effectively managing outliers with less severe penalties. This trait is beneficial for economic data prone to outliers from economic shocks or errors. MAE calculates the average error magnitude, offering straightforward interpretability without weighing the direction of errors. Its limitation is the equal treatment of all errors, potentially overlooking significant prediction mistakes. RMSE amplifies larger errors by squaring them, enhancing sensitivity to outliers, and spotlighting models with occasional inaccurate predictions, but this also means RMSE is less tolerant of outliers than MAE. SMAPE uses percentage errors, ensuring scale independence and suitability for comparing different datasets, although its reliability falters with values near zero, restricting its use in certain cases.

\bibitem{Note13}
The optimal hyperparameter configurations for our suite of machine learning models are as follows: For the Elastic Net model, the optimal penalty is 0.007739036873 and the mixture parameter is 0.6662955002. For the SVM (RBF kernel), the cost is 0.5971530254 and the RBF sigma is 0.003403106644. The k-NN model requires 11 neighbors, utilizing an inverse distance weighting function and a distance power of 0.9160159697. The Random Forest model has an mtry value of 39—indicating the number of features considered for splitting at each tree node—alongside a minimum node size of 12. For XGBoost, there are 861 trees and a learning rate of 0.01860805586, whereas the LightGBM model requires 1107 trees and a learning rate of 0.03441793697.

\bibitem{Note14}
Economies heavily reliant on agriculture experience pronounced fluctuations in overall output, exhibit low relative volatility in aggregate employment, and demonstrate a weak correlation between output and employment. Moreover, agriculture is distinctively characterized by more volatile output and employment, which are not positively correlated with those in other sectors, and a weaker correlation between output and employment within agriculture than in non-agricultural sectors. These unique characteristics significantly influence the dynamics of aggregate business cycles across countries~\cite {da2006role}.

\bibitem{Note2}
These studies encompass a wide range of topics, including the evolutionary dynamics of the international fossil energy trade's multilayer network~\cite {gao2015features}, the pivotal roles countries play within the international fossil fuel trade network~\cite {zhong2017roles}, the spatiotemporal evolution of global plastic waste trade networks~\cite {wang2020structure}, and the community structure of the food-trade international multi-network~\cite {torreggiani2018identifying}. Further contributions to the field examine the spatial-temporal evolution of the global copper raw materials and scrap trade networks~\cite {hu2020characteristics}, the characteristics and stability of the global complex nickel ore trade network~\cite {ma2022analysis}, the analysis of wood and non-wood forest global trade network~\cite {lovric2018social}, the temporal stability of international fertilizer trade networks, and the multiplex network structure of global cobalt industry chain~\cite {shi2022multiplex}.

\bibitem{Note3}
Prevailing studies on international trade networks predominantly utilize aggregate trade data or examine it at a highly granular level. Adopting a section-level perspective for trade data affords us a compromise between these two extremes, which is crucial since the granularity of data can significantly influence topological characteristics.

\bibitem{Note4}
Several works utilize complex networks in forecasting tasks, including the use of stock networks to predict economic growth~\cite {hegre2010trade} and the predictability of stock price series through community detection and network metrics~\cite {anghinoni2019time}.

\bibitem{Note5}
Refers to a comparative evaluation among a variety of machine learning models to determine which performs best on a given task or dataset. This analogy draws from actual horse races, where horses compete to see which is fastest under similar conditions to identify the top performer.

\bibitem{Note6}
We follow three-letter country codes defined in ISO 3166-1, part of the ISO 3166 standard published by the International Organization for Standardization (ISO).

\bibitem{Note7}
The Harmonized System of tariff nomenclatures, often referred to as the Harmonized System Classification, is a standardized system of names and numbers for classifying traded products. Developed and maintained by the World Customs Organization (WCO), the HS is utilized by over 200 countries and economies as a basis for their customs tariffs and for collecting international trade statistics. The HS is designed to facilitate the international trade process by simplifying the identification and classification of goods across borders. The system is structured into 21 sections, further divided into 97 chapters. These chapters are subdivided into headings and subheadings, representing a hierarchical structure that classifies goods from the general to the specific. For more information, access: \protect \href {https://www.wcoomd.org/en/topics/nomenclature/instrument-and-tools/hs-nomenclature-2022-edition/hs-nomenclature-2022-edition.aspx}{HS Nomenclature 2022 edition}.

\bibitem{Note8}
Originating in 1960, this index experienced its most pronounced increase during this timeframe. Specifically, the index's monthly average escalated from 27.3 (2010-2015) to 114.3 (2016-2019), with peaks exceeding 200 in certain months.

\bibitem{Note9}
GOSS (Gradient-based One-Side Sampling) gives preference to instances with larger gradients, indicating that they are more difficult to predict accurately, thereby focusing the learning process on these harder examples.

\bibitem{pearson1901liii}
K.~Pearson.
\newblock Liii. on lines and planes of closest fit to systems of points in space.
\newblock {\em The London, Edinburgh, and Dublin Philosophical Magazine and Journal of Science}, 2(11):559--572, 1901.

\bibitem{richardson2021nowcasting}
A.~Richardson, T.~van Florenstein~Mulder, and T.~Vehbi.
\newblock Nowcasting gdp using machine-learning algorithms: A real-time assessment.
\newblock {\em International Journal of Forecasting}, 37(2):941--948, 2021.

\bibitem{samitas2022covid}
A.~Samitas, E.~Kampouris, and S.~Polyzos.
\newblock Covid-19 pandemic and spillover effects in stock markets: A financial network approach.
\newblock {\em International Review of Financial Analysis}, 80:102005, 2022.

\bibitem{shi2022multiplex}
Q.~Shi, X.~Sun, M.~Xu, and M.~Wang.
\newblock The multiplex network structure of global cobalt industry chain.
\newblock {\em Resources Policy}, 76:102555, 2022.

\bibitem{silva2020micro}
T.~C. Silva, S.~M. Guerra, M.~A. da~Silva, and B.~M. Tabak.
\newblock Micro-level transmission of monetary policy shocks: The trading book channel.
\newblock {\em Journal of Economic Behavior \& Organization}, 179:279--298, 2020.

\bibitem{silva2017monitoring}
T.~C. Silva, S.~R.~S. Souza, and B.~M. Tabak.
\newblock Monitoring vulnerability and impact diffusion in financial networks.
\newblock {\em Journal of Economic Dynamics and Control}, 76:109--135, 2017.

\bibitem{silva2023effect}
T.~C. Silva, P.~V.~B. Wilhelm, and B.~M. Tabak.
\newblock The effect of interconnectivity on stock returns during the global financial crisis.
\newblock {\em The North American Journal of Economics and Finance}, 67:101940, 2023.

\bibitem{silva2016machine}
T.~C. Silva and L.~Zhao.
\newblock {\em Machine learning in complex networks}.
\newblock Springer, 2016.

\bibitem{smith2021identifying}
M.~Smith and F.~Alvarez.
\newblock Identifying mortality factors from machine learning using shapley values--a case of covid19.
\newblock {\em Expert Systems with Applications}, 176:114832, 2021.

\bibitem{torreggiani2018identifying}
S.~Torreggiani, G.~Mangioni, M.~J. Puma, and G.~Fagiolo.
\newblock Identifying the community structure of the food-trade international multi-network.
\newblock {\em Environmental Research Letters}, 13(5):054026, 2018.

\bibitem{vrontos2021modeling}
S.~D. Vrontos, J.~Galakis, and I.~D. Vrontos.
\newblock Modeling and predicting us recessions using machine learning techniques.
\newblock {\em International Journal of Forecasting}, 37(2):647--671, 2021.

\bibitem{wang2020structure}
C.~Wang, L.~Zhao, M.~K. Lim, W.-Q. Chen, and J.~W. Sutherland.
\newblock Structure of the global plastic waste trade network and the impact of china’s import ban.
\newblock {\em Resources, Conservation and Recycling}, 153:104591, 2020.

\bibitem{yule1897theory}
G.~U. Yule.
\newblock On the theory of correlation.
\newblock {\em Journal of the Royal Statistical Society}, 60(4):812--854, 1897.

\bibitem{zhong2017roles}
W.~Zhong, H.~An, L.~Shen, W.~Fang, X.~Gao, and D.~Dong.
\newblock The roles of countries in the international fossil fuel trade: An emergy and network analysis.
\newblock {\em Energy Policy}, 100:365--376, 2017.

\bibitem{zou2005regularization}
H.~Zou and T.~Hastie.
\newblock Regularization and variable selection via the elastic net.
\newblock {\em Journal of the Royal Statistical Society Series B: Statistical Methodology}, 67(2):301--320, 2005.

\end{thebibliography}

\end{document}